# High-efficiency magnetism modulation in a Weyl ferromagnet by current-assisted domain wall motion

Qiuyuan Wang[1,11,*], Yi Zeng[1,2,*], Kai Yuan[1,3], Qingqi Zeng[4], Pingfan Gu[1], Xiaolong Xu[1], Hanwen Wang[5], Zheng Han[6,7], Kentaro Nomura[8], Wenhong Wang[4,9], Enke Liu[4,9,§], Yanglong Hou[2,§], Yu Ye[1,3,10,§]

[1] State Key Laboratory for Mesoscopic Physics and Frontiers Science Centre for Nano-optoelectronics, School of Physics, Peking University, Beijing 100871, China
[2] Beijing Key Laboratory of Magnetoelectric Materials and Devices, Beijing Innovation Centre for Engineering Science and Advanced Technology, School of Materials Science and Engineering, Peking University, Beijing 100871, China
[3] Collaborative Innovation Centre of Quantum Matter, Beijing 100871, China
[4] Beijing National Centre for Condensed Matter Physics and Institute of Physics, Chinese Academy of Sciences, Beijing 100190, China
[5] Shenyang National Laboratory for Materials Science, Institute of Metal Research, Chinese Academy of Sciences, Shenyang 110016, China
[6] State Key Laboratory of Quantum Optics and Quantum Optics Devices, Institute of Opto-Electronics, Shanxi University, Taiyuan 030006, China
[7] Collaborative Innovation Center of Extreme Optics, Shanxi University, Taiyuan 030006, China
[8] Institute for Materials Research and Centre for Spintronics Research Network, Tohoku University, Sendai 980-8577, Japan
[9] Songshan Lake Materials Laboratory, Dongguan 523808, Guangdong, China
[10] Peking University Yangtze Delta Institute of Optoelectronics, Nantong 226010, Jiangsu, China
[11] Department of Electrical Engineering and Computer Science, Massachusetts Institute of Technology, Cambridge, Massachusetts 02139, USA
[*] These authors contributed equally.
[§] E-mail: ekliu@iphy.ac.cn, hou@pku.edu.cn, ye_yu@pku.edu.cn

**Flexible and efficient manipulation of magnetic configurations can be challenging. In the design of practical devices, achieving high effective magnetic field with low working current is under tight demand. Here, we report a unique method for efficient magnetism modulation by direct current injection in magnetic Weyl semimetal $Co_3Sn_2S_2$. We demonstrate that the modulation process stems from current-assisted domain wall motion. Through two independent methods, we reveal that the spin-transfer torque efficiency of $Co_3Sn_2S_2$ reaches as high as 2.4-5.6 kOe $MA^{-1}$ $cm^2$, and the threshold current density for driving the magnetic domain walls is as low as < $5.1\times10^5$ A/$cm^2$ without an external field, and < $1.5\times10^5$ A/$cm^2$ with a moderate external field. Our findings manifest a new and powerful approach for sub-micron magnetism manipulation, and also open the door towards a new paradigm of spintronics that combines magnetism, topology, and metallicity for low-energy consumption memory and computing.**

The interaction between current and magnetic domain wall (DW) is of fundamental interest in spintronics and has inspired a variety of applications in memory and logic devices, such as racetrack memory, to show unprecedented results[1–4]. When a spin-polarized current passes through a non-uniform magnetic texture such as a magnetic DW, spin-transfer torques (STTs) are applied to the local magnetization, so that the magnetic configuration can be directly controlled without the need to use the multilayer film structure required in the spin-orbit torque (SOT) approach[5–13]. But the STT approach also has its own problems. First, the use of STT to push DWs in metallic systems requires a current density threshold of $10^6$-$10^8$ A/$cm^2$ due to the intrinsic and extrinsic pinning effects, which increases the energy consumption of the device and limits its application[14–17]. Although a lower threshold current density can be realized at the cost of a lower DW velocity or a smaller depinning field in a semiconductor[18,19], an efficient approach that overcomes the magnetic DW pinning to achieve higher DW velocities in metallic systems is under tight demand for practical applications. Second, few attempts have been made to modulate the magnetic properties such as the coercive field in a single material layer utilizing STT, while SOT can be used in some cases when (local) inversion symmetry is broken[20,21]. One reason is that it is challenging to generate non-uniform magnetic textures in a uniformly magnetized sample by STT alone.

Here we report that in the synthesized $Co_3Sn_2S_2$ flakes (35-165 nm thick, grown on sapphire or $SiO_2$ substrates), a low current density on the order of $10^5$-$10^7$ A/cm$^2$ can reduce the coercive field $H_C$ of $Co_3Sn_2S_2$ from as high as 20 kOe down to smaller than 0.1 kOe. The threshold current to reduce $H_C$ as well as the symmetry of the +/– current can be further tuned by the geometry of the contact electrode. We reasonably deduce that the magnetism modulation behavior mainly comes from the current-assisted DW motion, specifically, through STT. To evaluate the STT efficiency, we fabricated $Co_3Sn_2S_2$ nanowire devices and measured their DW velocities in the parameter space of current, magnetic field, and temperature. We find an impressively large STT effective field of 2.4-5.6 kOe MA$^{-1}$ cm$^2$ at 150 K, which is even larger at lower temperatures, consistent with the results obtained by the reduction of coercive field in Hall devices. Also, very low current-induced DW motion thresholds of $< 5.1\times10^5$ A/cm$^2$ (without external field) and $< 1.5\times10^5$ A/cm$^2$ (with modest external field) in $Co_3Sn_2S_2$ are obtained. These findings demonstrate the high efficiency of the current-assisted DW motion and thus magnetism modulation in $Co_3Sn_2S_2$.

**Magnetism modulation with ultra-high efficiency in $Co_3Sn_2S_2$**

$Co_3Sn_2S_2$ is a ferromagnetic Weyl semimetal with a quasi-2D crystal structure consisting of stacked kagome lattices of cobalt atoms (Fig. 1a) and three pairs of Weyl points that are ~50 meV above the Fermi level in the first Brillouin zone[22–25]. In our work, high-quality single-crystalline $Co_3Sn_2S_2$ nanoflakes were directly grown on substrates (sapphire or $SiO_2$) via a modified chemical vapor transport method and then fabricated into standard Hall-bar devices (Fig. 1b, see details in Fig. S1 and Methods). From the electrical transport measurements, the $Co_3Sn_2S_2$ nanoflake shows ferromagnetism with a Curie temperature of about 175 K (Fig. 1c) and a large anomalous Hall angle up to 25.0%, consistent with those of reported bulk $Co_3Sn_2S_2$ single crystals[22,23,26].

The magnetization of $Co_3Sn_2S_2$ lies perpendicular to the sample plane. Due to the low density of states at Fermi energy and the sensitivity of the electronic band structure to magnetization direction, $Co_3Sn_2S_2$ exhibits a relatively small saturation magnetization $M_S$ of 29.3-48.8 emu/cm$^3$ as an itinerant ferromagnet, but an extremely large out-of-plane anisotropy field $H_K$ of 80-118 kOe at 150-170 K (the temperature range of this paper's main focus), which is much larger than the coercive field $H_C$ of no larger than 4 kOe in this temperature range[27]. The corresponding magnetic anisotropy constant $K$ is $(1.17\text{-}2.88)\times10^6$ erg/cm$^3$ (full description of electrical and magnetic properties in Supplementary Note 2). In both SOT and STT cases, small magnetization will result in high current-induced torque efficiency, which is expected to be inversely proportional to the free layer magnetization. For example, in the flow (i.e., linear) region, the current-induced DW velocity is $v_{\mathrm{DW}} \propto PJ/M_{\mathrm{S}}$ ($P$, the spin polarization of the current; $J$, the current density). Meanwhile, the field-driven DW velocity is $v_{\mathrm{DW}} \propto \Delta = (A/K)^{1/2}$ ($\Delta$, magnetic DW width constant; $A$, exchange stiffness). In other words, the large magnetic anisotropy will not only form thinner DWs, allowing downscaling of spintronic devices, but also reduce the perturbation from external fields, improve device stability, and further increase the current-induced effective field[28–30].

To explore the significant current-induced effects in $Co_3Sn_2S_2$, we measured the anomalous Hall resistance, $R_{yx}$, hysteresis loops of the Hall devices under different DC currents. Surprisingly, in this simple device structure containing only one material – $Co_3Sn_2S_2$ and Cr/Au contacts, a clear current modulation pattern of the hysteresis loop is observed (Fig. 1d and Fig. 1e from device 1). When the applied DC current is lower than the threshold current, the hysteresis loop and coercive field remain roughly unaffected. But when the current exceeds the threshold, the coercivity begins to drop rapidly, and eventually reaches a remarkably small value as the current increases. The observed change of $H_C$ deviates significantly from the value estimated by pure thermal effect (Fig. 1d). The current threshold behaviors are observed at all measured temperatures from 5 K to 150 K (more data provided in Supplementary Fig. S3). Eight samples from three different growth batches are tested, and all samples show the same current modulation behavior. These devices have a thickness of 35-165 nm, a width of 4-6 μm, and a length of 10-40 μm, and they are grown and fabricated on sapphire or $SiO_2$ substrates, demonstrating the robustness of this phenomenon. At 5 K, the coercive field does not decrease to the lowest value (Fig. 1d) because we limit the maximum applied current to 5.4 mA to protect the device for subsequent measurements. In another device, $H_C$ up to 20 kOe is tuned to < 0.1 kOe by current (Supplementary Fig. S4). Noticeably, if we define the efficiency of the current-induced magnetism modulation effect as the slope $\Delta H_C/\Delta j$ in the region above the threshold (thermal effect contribution is subtracted, see detailed discussion of thermal effect in Supplementary Note 4), it reaches 3.0 kOe $MA^{-1}$ $cm^2$ (150 K) - 10.5 kOe $MA^{-1}$ $cm^2$ (50 K). For comparison, the effective field generated by SOT in BiSb/MnGa ($\theta_{SH} \approx 52$) is 2.3 kOe $MA^{-1}$ $cm^2$, while the reported STT efficiency of driving magnetic DWs is 0.4 kOe $MA^{-1}$ $cm^2$ in (Ga,Mn)(As,P) and only 0.2-8.0 Oe $MA^{-1}$ $cm^2$ in Co-Pt-based systems[12,31,32].

**Current-assisted domain wall motion model**

The high efficiency of current-induced $H_C$ reduction is unusual. To unravel the underlying mechanism, we first investigated the magnetization reversal nature of the $Co_3Sn_2S_2$ samples. Figure 2a shows the hysteresis loop evolution as a function of the field angle in the *xz*-plane. $H_C$ increases monotonically with $\theta$ and follows the standard 1/cos$\theta$ behavior (Fig. 2b), indicating that the magnetization reversal process in $Co_3Sn_2S_2$ is dominated by DW motion rather than coherent rotation[33,34]. The multi-domain nature of the switching is also evidenced by the fact that $H_C \ll H_K$ (Fig. S2d). As a result, any macro-spin model fails. Meantime, all hysteresis loops are rectangle-shaped, clarifying that the mobile DWs suddenly appear at $H_C$ (nucleation) and then continue to move (propagation) until the magnetization of the whole sample is reversed.

Next, devices with asymmetric contact electrodes (devices 2 and 3) were fabricated to explore how current affects the DW nucleation and propagation. Figure 2c shows the representative hysteresis loops of device 2 under different currents, and Fig. 2d plots the variation of $H_C$ with the current. In this device, several characteristic features imply that the DW nucleation process is greatly modified by DC current injection, therefore the magnetic hysteresis loops and $H_C$ are modulated. First, under the small current when $H_C$ begins to decrease, the hysteresis loop remains rectangular (–7.6 mA in Fig. 2c) and shrinks with the current, indicating that the coercive field $H_C$ is still determined by the nucleation field $H_n$. Under larger currents, the hysteresis loop becomes irregular, and the sample enters a multi-domain state (–11.1 mA and 4.6 mA in Fig. 2c). In this region, $H_n$ is the field when the total magnetization starts to drop, while DW propagation determines the field that fully reverses the sample. As shown in Fig. 2c, both the nucleation and propagation fields are reduced by large currents. Second, an abnormal increase of $H_C$ with the increase of the current absolute value occurs between –(0.9-1.7) mA. Marked by the red dashed line in Fig. 2d, starting at the same magnitude of the positive current threshold $|I^+_{C,1}|$=$|I^-_{C,1}|$=0.9 mA, $H_C$ decreases with $|I^+_{C,1}|$, but increases with $|I^-_{C,1}|$, which shows that $\Delta|H_C|$=$\Delta|H_n|$ is an odd function of current. This symmetry is different from the common current effects in perpendicularly magnetized samples (details in Supplementary Note 5), including Joule heating (even function of current, regardless of homogenous or nonhomogeneous cases), SOT (even function of current when the external field is out-of-plane) and Oersted field (odd function of the field)[35,36]. There must be some other responsible effects, which we will attribute to the current-assisted DW motion that altered the DW nucleation process discussed later in this paper. Third, the appearance of another platform of $H_C$ and a larger negative current threshold $I^{-}_{C,2} \approx -6.2$ mA (the blue dashed line in Fig. 2d) indicates that multiple nucleation sites are involved, otherwise this non-monotonic change on the negative current side will not occur.

To further study the contributive nucleation location of DW and clarify its STT origin, we designed a device (device 3) with two drain electrodes of different shapes and contact areas (Fig. 2e). When current is injected through paths 1-2 or 1-3, the path with the smaller electrode 2 shows an evidently smaller positive current threshold than that of electrode 3, while the negative current thresholds of the two paths are almost the same (Fig. 2f). This signifies that the effective nucleation of DW is close to the source or drain electrodes – the negative current only contributes to the DW nucleation near electrode 1, while the positive current only contributes to the DW nucleation near electrode 2 or 3. The local and unidirectional nature of this interaction further suggests the STT origin of this phenomenon. Besides, the unidirectionality confirms that the direction of the current-induced force on magnetic DWs is along the electron flow direction (consistent with the DW velocity measurement results discussed later), and demonstrates that the current modulation behavior can be artificially controlled by the design of the electrode geometry.

Now we can use the current-assisted DW motion model to describe all the observed magnetism modulation phenomena (Fig. 2g). As mentioned earlier, under a moderate current, the coercive field $H_C$ of the sample is determined by the nucleation field $H_n$. Note that there can be multiple nucleation sites, and $H_C$ equals the lowest nucleation field, that is, $H_C = H_{n,\min}$. From the above experiments, the current will affect the DW nucleation process near the source and drain electrodes. We assume that the injected current applies STT on the DW. Since the force induced by STT is an odd function of current, its influence on $H_n$ depends on the current direction. As the magnetization easy-axis is out-of-plane and perpendicular to the current, the magnetization direction will not affect $H_n$. For a specific nucleation location, $H_n$ decreases (increases) when the current induced force is along (against) the DW propagation direction. Figure 2g illustrates how the DC current affects the DW nucleation process at two different locations with opposite DW propagation directions. The blue and red lines represent the $H_n$ at the two locations, whose nucleation fields and current thresholds are slightly different. Then, the coercive field $H_C$ is represented by the solid lines, which is determined by the minimum $H_n$ values under different applied currents. We found that this simple model captured the two main abnormal features we observed experimentally in Fig. 2d. First, the decisive nucleation position is switched at $H_{n,1} = H_{n,2}$ (marked by the purple arrows in Fig. 2d and Fig. 2g), resulting in asymmetric threshold currents of $I^{+}_{C,1} \approx 1.2$ mA and $I^{-}_{C,2} \approx -6.2$ mA. Second, $H_C$ increases abnormally with the current at $I^{-}_{C,1}$ (symmetric to $I^{+}_{C,1}$), as expected in this current-assisted DW motion picture. This model is further verified by the control experiments illustrated in Fig. 3c-h, and a detailed discussion of the effect of current on nucleation can be found in the Supplementary Information Note 5 and 7.

**Domain wall velocity measurement revealing high STT efficiency**

Despite the above model qualitatively explains the magnetism modulation phenomenon, it lacks a quantitative measurement of DW dynamics. To better assign the current-assisted DW motion model and understand its ultra-high efficiency, we managed to measure the field and current dependent DW velocity in $Co_3Sn_2S_2$ through the time-of-flight method[37,38]. The measurement scheme is depicted in Fig. 3a. We fabricated a 1 μm wide and 62 nm thick $Co_3Sn_2S_2$ nanowire with three pairs of Hall bars in the middle, two Au heaters on both sides, and source/drain electrodes (device 4). The sample magnetization is first saturated by a negative field larger than $H_C$. Next, the field is changed to positive $H_z < H_C$, and a DC current $I_{DC}$ is applied in the nanowire. At this time, the sample is magnetized in the $-z$-direction. The magnetic DW is then prepared by injecting a current pulse into the heater (the left one in the schematic). Under the mutual effect of current, field and heating, a DW is generated near the heater and propagates in the opposite direction of $I_{DC}$ (from left to right in the schematic). The motion of the DW is monitored by an oscilloscope recording the Hall voltage change (thus the change of $R_{yx}$) on two Hall bars separated by 15 μm, so that the DW velocity can be obtained as a function of $H_z$ or $I_{DC}$. The typical waveforms of the heater pulse and $R_{yx}$ changes are shown in Fig. 3b. Clearly, the DW propagates across Hall bar 2 and 3 subsequently (along with the direction of the electron flow), resulting in a $2R_0$ change in $R_{yx}$, where $R_0$ is the saturated anomalous Hall resistance. We also used a lock-in amplifier to monitor the $R_{yx}$ at Hall bar 1 to ensure the magnetization direction and saturation of the nanowire before and after the current pulse.

Before the DW dynamics measurement, we utilized the nanowire device to cross-check the role of Joule heating and current-assisted DW motion by STT in the magnetism modulation process (Fig. 3c-h). A tiny 1 μA AC current is used to track the Hall resistance $R_{yx}$, while the effects of Joule heating and STT are separated by the DC current injection into the heater and $Co_3Sn_2S_2$ nanowire, respectively. Figures 3c-d show the $R_{yx}$ hysteresis loops as a function of the heater current $I_{heater}$ from the configuration that current is only applied on the heater. When $I_{heater}$ is large enough, the temperature near the heater will rise, lowering the energy barrier for DW nucleation, so $H_C$ will decrease. Because it is a pure thermal effect, negative $I_{heater}$ also has the same effect (not shown in the figure). In addition, $I_{heater}$ will not increase $H_C$. Figures 3e-f show the results when both the heater and the nanowire run current. $I_{heater}$ is now fixed at 15 mA to release the magnetic DW near the heater and let it propagate from left to right during the magnetic reversal process. The resulting hysteresis loops as a function of the current in the nanowire $I_{sd}$ (positive direction defined from right to left) are shown in Fig. 3f. Contrary to the heater-only case (Fig. 3d), +/– $I_{sd}$ show opposite modulation behavior to $H_C$. Positive $I_{sd}$ exerts a force on the DW away from the heater to help its nucleation and propagation process, thus reducing $H_C$, and vice versa. In the intermediate current region between –0.2 mA and 0 mA, the modulation efficiency $\Delta H_C/\Delta J$ reaches 2.2 kOe $MA^{-1}$ $cm^2$, very close to the value obtained in device 1. In Fig. 3g-h, when only $I_{sd}$ is applied, the hysteresis loop remains unchanged in the same $I_{sd}$ range as in Fig. 3e-f, indicating that $I_{sd}$ does not bring severe heating effect. The comparison of the three cases in Fig. 3c-h illustrates that STT plays an important role in the magnetism modulation process by affecting DW nucleation, thus resulting in the ultra-high efficiency, while Joule heating ensures that the dominated nucleation process occurs near the source/drain electrodes, so DW propagates unidirectionally during the magnetization reversal process and $H_n$ can be effectively modulated.

We finally perform DW velocity measurements to give an unambiguous result of STT efficiency. Figure 3i shows the contour plot of the DW velocity $v_{DW}$ as a function of $H_z$ and $J$, where three "iso-speed" lines with an interval of 1 m/s are displayed by the black dashed lines. Interestingly, $v_{DW}$ shows an excellent linear dependence on the field and current density covering our whole measurement range, which can be expressed as $v_{DW} = \eta J + \mu H_z$, where $\eta$ and $\mu$ are the current- and field-dependent domain wall mobility, respectively. The linear relationship can be seen more clearly in Fig. 3j,k, where the dependence of $v_{DW}$ on $J$ and $H_z$ is measured at various temperatures, showing that the magnetic DW enters the linear flow region under a small current density of ~ $10^5$ A/cm$^2$ and magnetic field of ~ 0.1 kOe [28]. The linear relationship demonstrates the field-like effect of STT on DW dynamics, so the effective field of STT can be obtained as $\eta/\mu$, reaching 2.4 kOe $MA^{-1}$ $cm^2$ at 150 K in device 4, and 5.6 kOe $MA^{-1}$ $cm^2$ in another device (device 5, Supplementary Fig. S9). Note that the magnetism modulation efficiencies observed in devices 1-3 at the same temperature are 3.0-3.8 kOe $MA^{-1}$ $cm^2$, which is within the STT efficiency range obtained by DW dynamics measurements. A full comparison between these two efficiencies obtained from different devices is depicted in Fig. 4. Their values and the temperature-dependences are well matched (see Supplementary Note 7), which strengthens the current-assisted DW motion model and provides direct evidence of the ultra-high STT efficiency in $Co_3Sn_2S_2$.

Another crucial parameter in DW dynamics is the threshold current density $J_c$, above which the current-induced DW motion can be achieved. In our experiments, even the lowest applied current density exceeds the threshold because a finite magnetic field and current are required to nucleate the DW and generate observable electrical signals. As a result, only the upper bound of $J_c$ can be given. We further calculate an effective actual threshold value $J_c^* = J_c + (\mu/\eta) H_z$, which is ensured by the field-like effect of the current. From Fig. 3j, we see that $J_c < 1.6\times10^5$ A/cm$^2$ under 0.2 kOe field, giving $J_c^* < 2.8\times10^5$ A/cm$^2$ at 167.5 K. A lower threshold is observed in device 5, with $J_c < 1.0\times10^5$ A/cm$^2$ under 0.2 kOe field and $J_c^* < 1.5\times10^5$ A/cm$^2$ at 160 K. In addition, DW nucleation is occasionally achieved without magnetic field in this device. Therefore, we obtain the maximum upper bound of $J_c < 5.1\times10^5$ A/cm$^2$ at 160 K (Supplementary Fig. S9), on the same order of magnitude as $J_c^*$, which is still two orders of magnitude smaller than the value of common metallic ferromagnets.

## Discussion

In conclusion, we find that DC current significantly modulates the coercivity and thus magnetization reversal process in $Co_3Sn_2S_2$ nanoflakes with ultra-high efficiency. The mechanism is carefully studied and confirmed to be current-assisted DW motion. Current- and field-driven DW velocity measurement further reveals the very large STT efficiency and low threshold current density in $Co_3Sn_2S_2$. Remarkably, these unprecedented parameters are fully confirmed by analytical DW motion model and micromagnetic simulation (details in Supplementary Note 8), suggesting that the direct cause of the unusual phenomena is classical magnetic and transport properties: large $K_u$, small $M_S$, and large $P$. Nevertheless, the combination of these benefits is closely related to the special electronic band structure of a magnetic Weyl semimetal, and hard to achieve in conventional ferromagnetic metals, making magnetic Weyl semimetals an attractive material category for energy-efficient spintronic devices. This work provides a new scheme of magnetism manipulation, demonstrates that STT in metallic systems can produce a large effective field comparable to the best results of SOT. Direct control of magnetism by current injection into a single material with large perpendicular magnetic anisotropy may enable more flexible design on DW functional devices, which are more compact, stable, and power-saving.

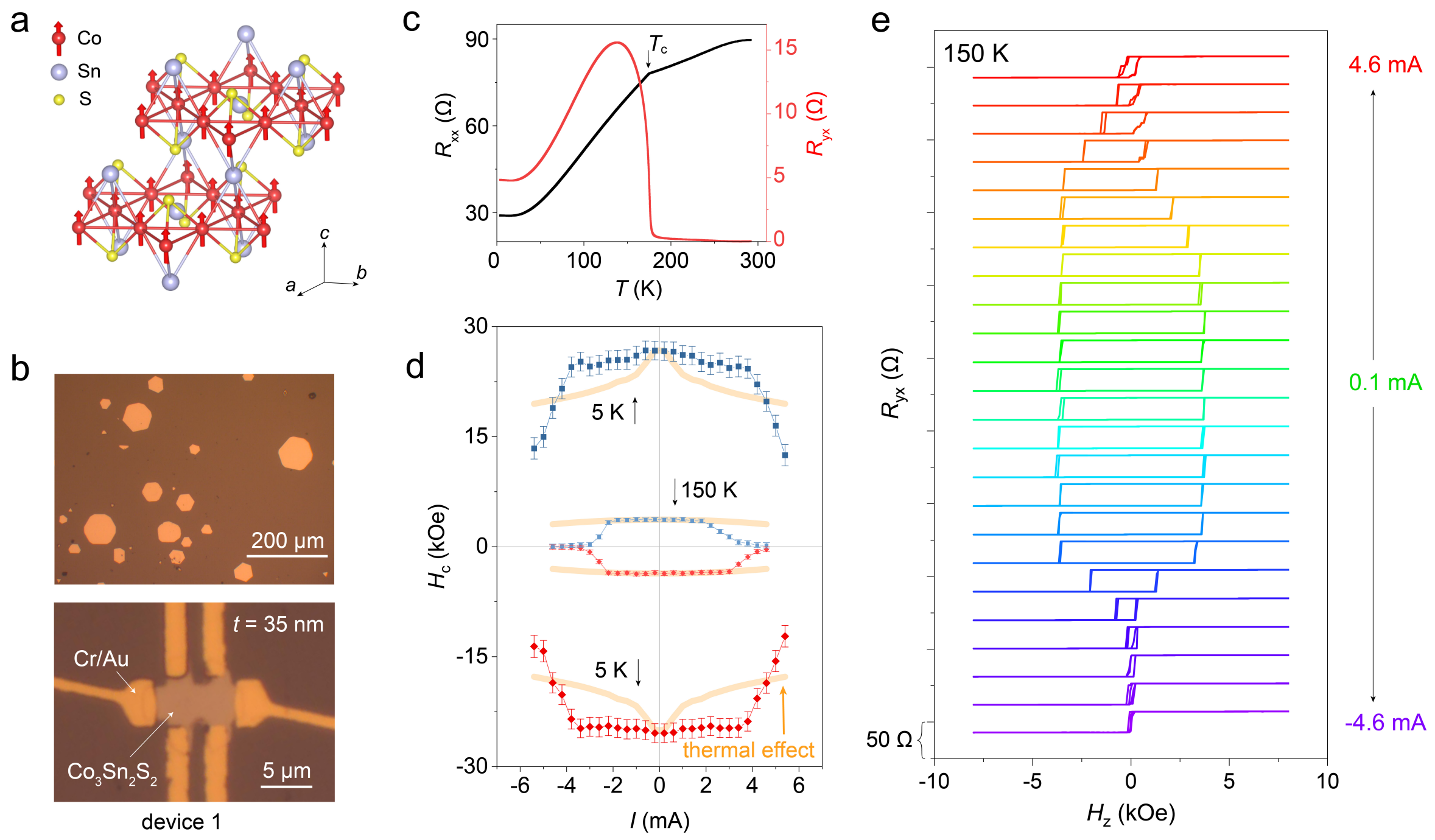

**Fig. 1 Current modulation of the reversal process. a,** Crystal structure of $Co_3Sn_2S_2$ with a space group of *R-3m* (no. 166). The cobalt atoms form a quasi-2D kagome lattice. **b,** Optical images of the as-grown $Co_3Sn_2S_2$ nanoflakes on a sapphire substrate and a fabricated 35 nm thick Hall-bar device (device 1). **c,** The temperature dependence of the longitudinal ($R_{xx}$) and transverse ($R_{yx}$) resistance under 0.1 kOe field cooling along the *c*-axis. A Curie temperature of $T_C \approx 175$ K is obtained. **d,** When the applied DC current exceeds the threshold, the coercive field $H_C$ (red and blue points) begins to decrease rapidly, showing significant discrepancy from the pure thermal effect (orange lines, obtained by using the value of $R_{yx}$ as an internal thermometer). **e,** The hysteresis loops of $R_{yx}$ under different DC currents measured at 150 K. Each loop is measured twice to ensure repeatability.

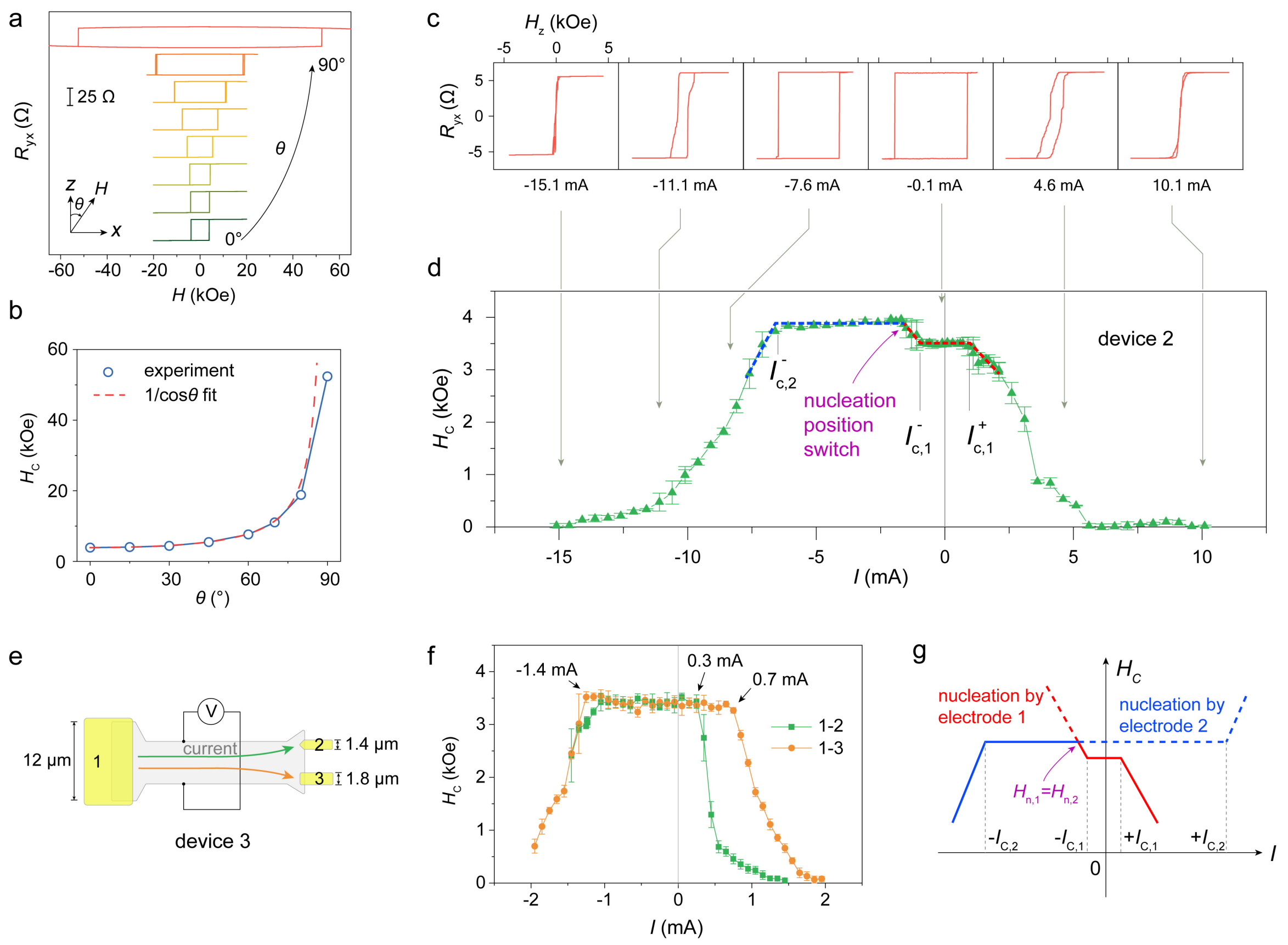

**Fig. 2 Current-assisted domain wall motion model. a,** Field angle dependence of $R_{yx}(H)$. The sample lies in the *xy*-plane. Magnetic field $H$ is applied in the *xz*-plane, and $\theta$ is the angle between $H$ and the *z*-axis. **b,** Coercive field $H_C$ as a function of $\theta$. Error bars (both horizontal and vertical) are smaller than data points. The red dashed line represents $H_C(\theta=0)/\cos\theta$, the predicted value when the magnetization reversal process is dominated by DW motion. **c-d,** $H_C$ and representative hysteresis loops of device 2 (40 μm×6 μm×92 nm) under different DC currents, showing features including the abnormal increase of $H_C$, asymmetric current thresholds, and multi-domain. **e,** Schematic of device 3 (25 μm×5.5 μm×165 nm), with one large electrode at one side and two small electrodes on the other side. Electrodes 2 and 3 are triangular and rectangular shaped, respectively. **f,** $H_C$ as a function of DC current and the current path in device 3. **g,** The evolution of $H_C$ given by the current-assisted DW motion model in the simple case where two nucleation sites with different DW nucleation field $H_n$ located under source and drain electrodes. $H_n$ at the two sites are colored red and blue, respectively, and the smallest $H_n$ determines the coercive field (solid line). When the current reaches the value marked by the purple arrow, where $H_{n,1}=H_{n,2}$, the decisive nucleation site changes (also shown in **d**). All data in **Fig. 2** are measured at 150 K.

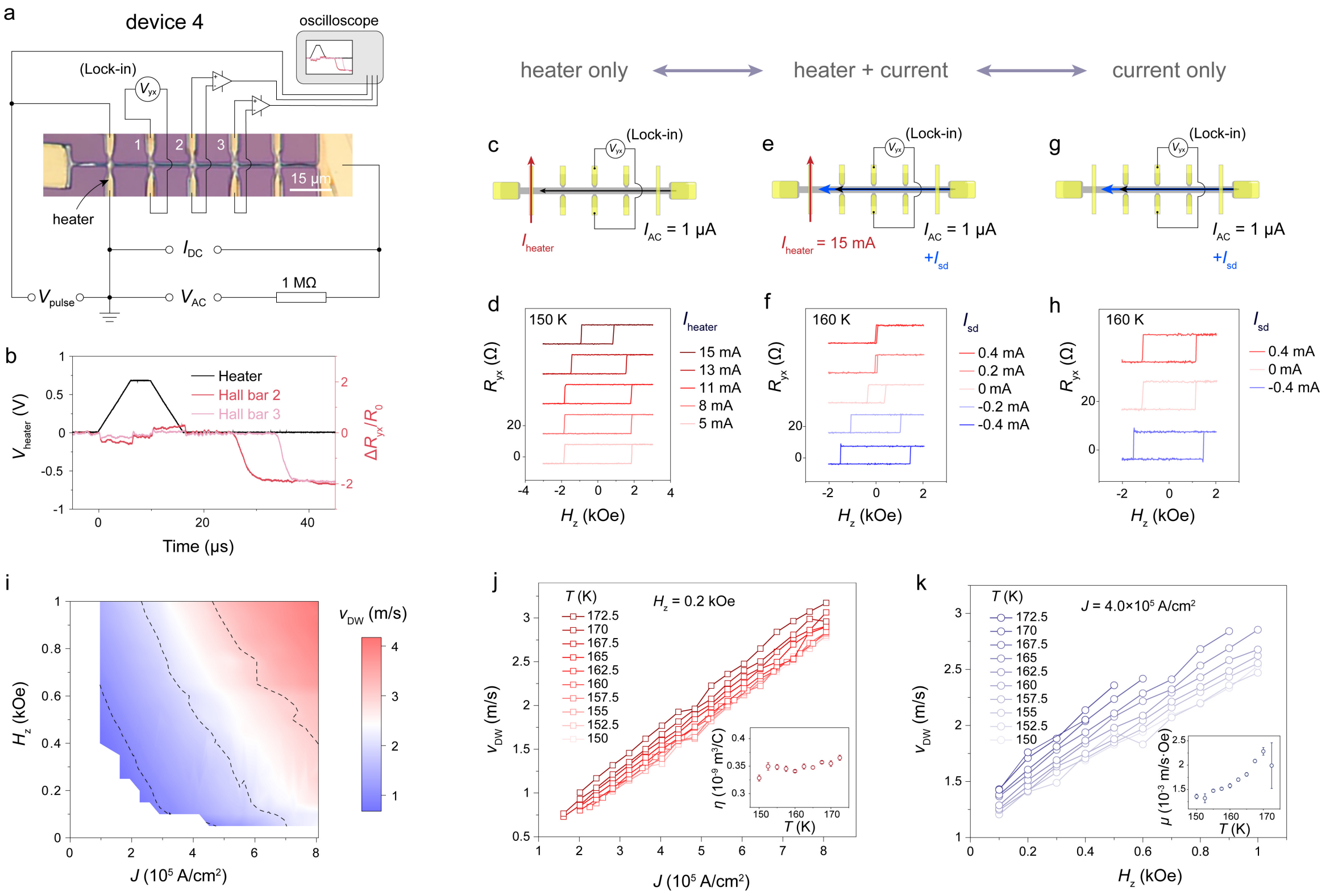

**Fig. 3 Domain wall velocity measurement. a,** Schematic diagram of the domain wall velocity measurement and optical image of the tested nanowire device (device 4, 90 μm×1 μm×62 nm). The Hall voltage signals are transmitted to the oscilloscope through 1000× gain preamplifiers. **b,** Typical waveform obtained during a measurement. Black line shows the voltage pulse applied onto the heater. Red lines trace the change of $R_{yx}$. Domain wall velocity $v_{DW}$ equals the channel length between the two Hall bars divided by the time interval of $R_{yx}$ reversal. **c-h,** Control experiments to separate the contribution of Joule heating (by $I_{heater}$) and STT (by $I_{sd}$) in the magnetism modulation process. The $R_{yx}$ hysteresis loops are measured through 1 μA AC current that minimizes its influence on magnetization reversal. **i,** Contour map of $v_{DW}$ as a function of out-of-plane field $H_z$ and current density $J$. Black dashed lines show "iso-speed" lines with a spacing of 1 m/s. **j,** $v_{DW}$ as a function of $J$ and $T$, measured under $H_z = 0.2$ kOe. Inset, the current-dependent DW mobility $\eta$ as a function of $T$. **k,** $v_{DW}$ as a function of $H_z$ and $T$, measured at $J = 4.0\times10^5$ A/cm$^2$. Inset, the field-dependent DW mobility $\mu$ as a function of $T$.

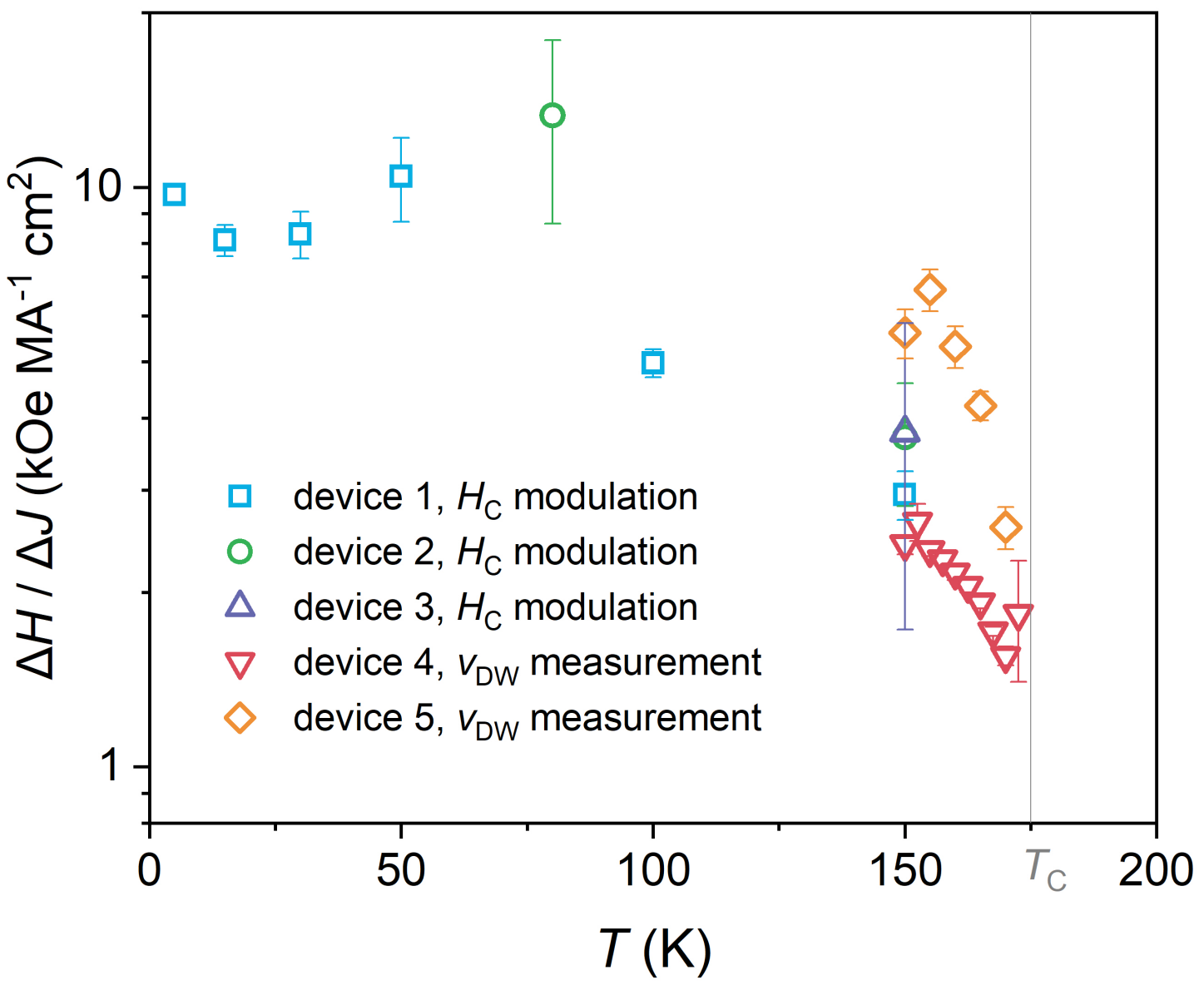

**Fig. 4 The STT efficiency $\Delta H/\Delta J$ given by two independent methods.** As described in the main text, devices 1-3 are Hall-bar devices, and the efficiency is given by the current modulation process of coercive field $H_C$, which is $\Delta H_C/\Delta J$. Device 4-5 are nanowire devices, and the efficiency is given by the current-induced effective field in DW velocity measurements, which is $\Delta H/\Delta J = \eta/\mu$ [$\eta(\mu)$, current(field)-dependent DW mobility]. These data from different devices and methods show good consistency, providing strong evidence of the large STT efficiency in $Co_3Sn_2S_2$.

## Methods

***Sample Growth and Device Fabrication*** $Co_3Sn_2S_2$ nanoflakes are grown on sapphire or $Si/SiO_2$ substrates via a modified chemical vapor transport method. $Co_3Sn_2S_2$ powders are sealed in one end of an ampule as precursors with transport agents, and substrates are placed at the other end (see Supplementary Fig. S1). To reduce the transport rate, a neck of a diameter of 1 mm is introduced in the middle of the ampule. The source and the sink ends are gradually heated to 1000 °C and 800 °C in 3 hours, and the temperatures are kept for 1 hour. After natural cooling to room temperature, high-quality crystals are obtained, confirmed by high resolution transmission electron microscope and selected area electron diffraction patterns. The thicknesses of the samples are measured by atomic force microscopy. The nanoflakes are patterned into designed geometry by electron beam lithography and argon ion milling. After that, Cr/Au (10 nm/100 nm) electrodes are defined by electron beam lithography and evaporation.

***Transport Measurement*** The device is loaded into a PPMS (Cryomagnetics, Inc.) with a magnetic field up to 14 T. For all-DC measurements (which is the case for devices 1-3), the current is applied by Keithley 2636B. Longitudinal and transverse voltage are measured by Keithley 2636B and 2182A, respectively. The devices show linear *I*-*V* curves in the measured current range, so the resistance can be directly obtained by dividing the voltage by the DC current. A small current of 100 μA is used to obtain the *R*-*T* and $H_C$-$\theta$ data. A self-made rotator probe is used for angular-dependent measurements. For current modulation measurements, the hysteresis loops can be obtained by directly sweeping the magnetic field or by holding the field but rotating the sample.

***Domain Wall Velocity Measurement*** The schematic diagram to measure the DW velocity is shown in Fig. 3a. The AC signal is applied and measured by SR830 lock-in amplifier. The DC current is applied by Keithley 2636B. The current pulse is generated by AFG31000 arbitrary function generator. The Hall voltage signal is magnified 1000× by Signal Recovery 5186 preamplifier and is then traced by DPO2024B oscilloscope. The pulse duration on the heater is set to 10 μs to ensure DW generation, with rise/fall time set to be 5 μs to minimize interference on the Hall signal. As a result, clean jump signals are obtained when DW passes the Hall bar. Background signals are filtered by a 0.5 Hz cut-off frequency of the preamplifier.

**Data availability**
The datasets generated by the present study are available from the corresponding author upon request.

**Acknowledgements**
The authors thank L. Liu and Y. Zhu for valuable discussions, X. Cheng for RMCD measurements, and J. Zhang for MFM measurements. This work was supported by the National Key R&D Program of China (Grant Nos. 2017YFA0206301, 2018YFA0306900 and 2019YFA0704900), the National Natural Science Foundation of China (Grant Nos. 6152004, 11974394, 51631001 and 51672010), and the Strategic Priority Research Program (B) of the Chinese Academy of Sciences (XDB33000000).

**Author contributions**
Q.W. and Y.Y. conceived the project. Y.Z. grew the $Co_3Sn_2S_2$ nanoflakes and fabricated the devices assisted by K.Y.. Q.Z. and E.L. provided the precursor $Co_3Sn_2S_2$ powders and helpful discussion of preparing the nanoflake samples. Q.W. conducted the transport measurements assisted by Y.Z., K.Y., and P.G.. Q.W. analyzed the data. X.X. performed HRTEM measurements. E.L. and K.N. provided scientific discussions. Q.W. and Y.Y. wrote the manuscript, with input from all authors. Y.H. and Y.Y. supervised the project. All authors discussed the results, interpretation and conclusion.

**Competing interests**
The authors have filed a Chinese patent application for using direct current injection to modulate magnetism in magnetic semimetal thin films.

**References**

1. Myers, E. B. Current-Induced Switching of Domains in Magnetic Multilayer Devices. *Science* **285**, 867–870 (1999).
2. Parkin, S. S. P., Hayashi, M. & Thomas, L. Magnetic Domain-Wall Racetrack Memory. *Science* **320**, 190–194 (2008).
3. Luo, Z. *et al.* Current-driven magnetic domain-wall logic. *Nature* **579**, 214–218 (2020).
4. Raymenants, E. *et al.* Nanoscale domain wall devices with magnetic tunnel junction read and write. *Nat Electron* **4**, 392–398 (2021).
5. Berger, L. Exchange interaction between ferromagnetic domain wall and electric current in very thin metallic films. *Journal of Applied Physics* **55**, 1954–1956 (1984).
6. Tatara, G. & Kohno, H. Theory of Current-Driven Domain Wall Motion: Spin Transfer versus Momentum Transfer. *Phys. Rev. Lett.* **92**, 086601 (2004).
7. Zhang, S. & Li, Z. Roles of Nonequilibrium Conduction Electrons on the Magnetization Dynamics of Ferromagnets. *Phys. Rev. Lett.* **93**, 127204 (2004).
8. Tatara, G., Kohno, H. & Shibata, J. Microscopic approach to current-driven domain wall dynamics. *Physics Reports* **468**, 213–301 (2008).
9. Miron, I. M. *et al.* Perpendicular switching of a single ferromagnetic layer induced by in-plane current injection. *Nature* **476**, 189–193 (2011).
10. Liu, L., Lee, O. J., Gudmundsen, T. J., Ralph, D. C. & Buhrman, R. A. Current-Induced Switching of Perpendicularly Magnetized Magnetic Layers Using Spin Torque from the Spin Hall Effect. *Phys. Rev. Lett.* **109**, 096602 (2012).
11. Liu, L. *et al.* Spin-Torque Switching with the Giant Spin Hall Effect of Tantalum. *Science* **336**, 555–558 (2012).
12. Khang, N. H. D., Ueda, Y. & Hai, P. N. A conductive topological insulator with large spin

Hall effect for ultralow power spin–orbit torque switching. *Nature Mater* **17**, 808–813 (2018).
13. Dc, M. *et al.* Room-temperature high spin–orbit torque due to quantum confinement in sputtered BixSe(1–x) films. *Nature Mater* **17**, 800–807 (2018).
14. Yamaguchi, A. *et al.* Real-Space Observation of Current-Driven Domain Wall Motion in Submicron Magnetic Wires. *Phys. Rev. Lett.* **92**, 077205 (2004).
15. Feigenson, M., Reiner, J. W. & Klein, L. Efficient current-induced domain-wall displacement in SrRuO 3. *Physical review letters* **98**, 247204 (2007).
16. Koyama, T. *et al.* Current-induced magnetic domain wall motion below intrinsic threshold triggered by Walker breakdown. *Nature Nanotech* **7**, 635–639 (2012).
17. Gushi, T. *et al.* Large Current Driven Domain Wall Mobility and Gate Tuning of Coercivity in Ferrimagnetic $Mn_4N$ Thin Films. *Nano Lett.* **19**, 8716–8723 (2019).
18. Yamanouchi, M., Chiba, D., Matsukura, F. & Ohno, H. Current-induced domain-wall switching in a ferromagnetic semiconductor structure. *Nature* **428**, 539–542 (2004).
19. Yamanouchi, M. *et al.* Universality Classes for Domain Wall Motion in the Ferromagnetic Semiconductor (Ga,Mn)As. *Science* **317**, 1726–1729 (2007).
20. Liu, L. *et al.* Electrical switching of perpendicular magnetization in a single ferromagnetic layer. *Phys. Rev. B* **101**, 220402 (2020).
21. Zhang, K. *et al.* Gigantic Current Control of Coercive Field and Magnetic Memory Based on Nanometer-Thin Ferromagnetic van der Waals $Fe_3GeTe_2$. *Adv. Mater.* **33**, 2004110 (2021).
22. Liu, E. *et al.* Giant anomalous Hall effect in a ferromagnetic kagome-lattice semimetal. *Nature physics* **14**, 1125–1131 (2018).
23. Wang, Q. *et al.* Large intrinsic anomalous Hall effect in half-metallic ferromagnet Co3Sn2S2 with magnetic Weyl fermions. *Nat Commun* **9**, 3681 (2018).
24. Liu, D. F. *et al.* Magnetic Weyl semimetal phase in a Kagomé crystal. *Science* **365**, 1282–1285 (2019).
25. Morali, N. *et al.* Fermi-arc diversity on surface terminations of the magnetic Weyl semimetal $Co_3Sn_2S_2$. *Science* **365**, 1286–1291 (2019).
26. Tanaka, M. *et al.* Topological Kagome Magnet $Co_3Sn_2S_2$ Thin Flakes with High Electron Mobility and Large Anomalous Hall Effect. *Nano Lett.* **20**, 7476–7481 (2020).
27. Shen, J. *et al.* On the anisotropies of magnetization and electronic transport of magnetic Weyl semimetal $Co_3Sn_2S_2$. *Appl. Phys. Lett.* **115**, 212403 (2019).
28. Metaxas, P. J. *et al.* Creep and Flow Regimes of Magnetic Domain-Wall Motion in Ultrathin Pt / Co / Pt Films with Perpendicular Anisotropy. *Phys. Rev. Lett.* **99**, 217208 (2007).
29. Mougin, A., Cormier, M., Adam, J. P., Metaxas, P. J. & Ferré, J. Domain wall mobility, stability and Walker breakdown in magnetic nanowires. *Europhys. Lett.* **78**, 57007 (2007).
30. Beach, G. S. D., Tsoi, M. & Erskine, J. L. Current-induced domain wall motion. *Journal of Magnetism and Magnetic Materials* **320**, 1272–1281 (2008).
31. Díaz Pardo, R. *et al.* Common universal behavior of magnetic domain walls driven by spin-polarized electrical current and magnetic field. *Phys. Rev. B* **100**, 184420 (2019).
32. Lee, J.-C. *et al.* Universality Classes of Magnetic Domain Wall Motion. *Phys. Rev. Lett.* **107**, 067201 (2011).

33. A mechanism of magnetic hysteresis in heterogeneous alloys. *Phil. Trans. R. Soc. Lond. A* **240**, 599–642 (1948).
34. Cullity, B. D. & Graham, C. D. *Introduction to magnetic materials*. (2015).
35. Thiaville, A. & Nakatani, Y. Domain-Wall Dynamics in Nanowiresand Nanostrips. in *Spin Dynamics in Confined Magnetic Structures III* (eds. Hillebrands, B. & Thiaville, A.) vol. 101 161–205 (Springer Berlin Heidelberg, 2006).
36. Manchon, A. *et al.* Current-induced spin-orbit torques in ferromagnetic and antiferromagnetic systems. *Rev. Mod. Phys.* **91**, 035004 (2019).
37. Yoshimura, Y. *et al.* Soliton-like magnetic domain wall motion induced by the interfacial Dzyaloshinskii–Moriya interaction. *Nature Phys* **12**, 157–161 (2016).
38. Kim, K.-J. *et al.* Fast domain wall motion in the vicinity of the angular momentum compensation temperature of ferrimagnets. *Nature Mater* **16**, 1187–1192 (2017).

# Supplementary Information for

# High-efficiency magnetism modulation of $Co_3Sn_2S_2$ by current-assisted domain wall motion

Qiuyuan Wang[1,11,*], Yi Zeng[1,2,*], Kai Yuan[1,3], Qingqi Zeng[4], Pingfan Gu[1], Xiaolong Xu[1], Hanwen Wang[5], Zheng Han[6,7], Kentaro Nomura[8], Wenhong Wang[4,9], Enke Liu[4,9,§], Yanglong Hou[2,§], Yu Ye[1,3,10,§]

[1] State Key Laboratory for Mesoscopic Physics and Frontiers Science Centre for Nano-optoelectronics, School of Physics, Peking University, Beijing 100871, China
[2] Beijing Key Laboratory of Magnetoelectric Materials and Devices, Beijing Innovation Centre for Engineering Science and Advanced Technology, School of Materials Science and Engineering, Peking University, Beijing 100871, China
[3] Collaborative Innovation Centre of Quantum Matter, Beijing 100871, China
[4] Beijing National Centre for Condensed Matter Physics and Institute of Physics, Chinese Academy of Sciences, Beijing 100190, China
[5] Shenyang National Laboratory for Materials Science, Institute of Metal Research, Chinese Academy of Sciences, Shenyang 110016, China
[6] State Key Laboratory of Quantum Optics and Quantum Optics Devices, Institute of Opto-Electronics, Shanxi University, Taiyuan 030006, China
[7] Collaborative Innovation Center of Extreme Optics, Shanxi University, Taiyuan 030006, China
[8] Institute for Materials Research and Centre for Spintronics Research Network, Tohoku University, Sendai 980-8577, Japan
[9] Songshan Lake Materials Laboratory, Dongguan 523808, Guangdong, China
[10] Peking University Yangtze Delta Institute of Optoelectronics, Nantong 226010, Jiangsu, China
[11] Department of Electrical Engineering and Computer Science, Massachusetts Institute of Technology, Cambridge, Massachusetts 02139, USA
[*] These authors contributed equally.
[§] E-mail: ekliu@iphy.ac.cn, hou@pku.edu.cn, ye_yu@pku.edu.cn

**This PDF file includes:**
Supplementary Figures S1 to S10
Supplementary Tables S1 to S2
References

## Note 1: $Co_3Sn_2S_2$ nanoflake growth

The $Co_3Sn_2S_2$ nanoflakes can be grown through chemical vapor transport (CVT) or co-sputtering method[1–4]. Here, we used a modified CVT method to grow high-quality single-crystalline $Co_3Sn_2S_2$ nanoflakes, with lateral size up to ~100 μm and thickness of several tens of nanometers.

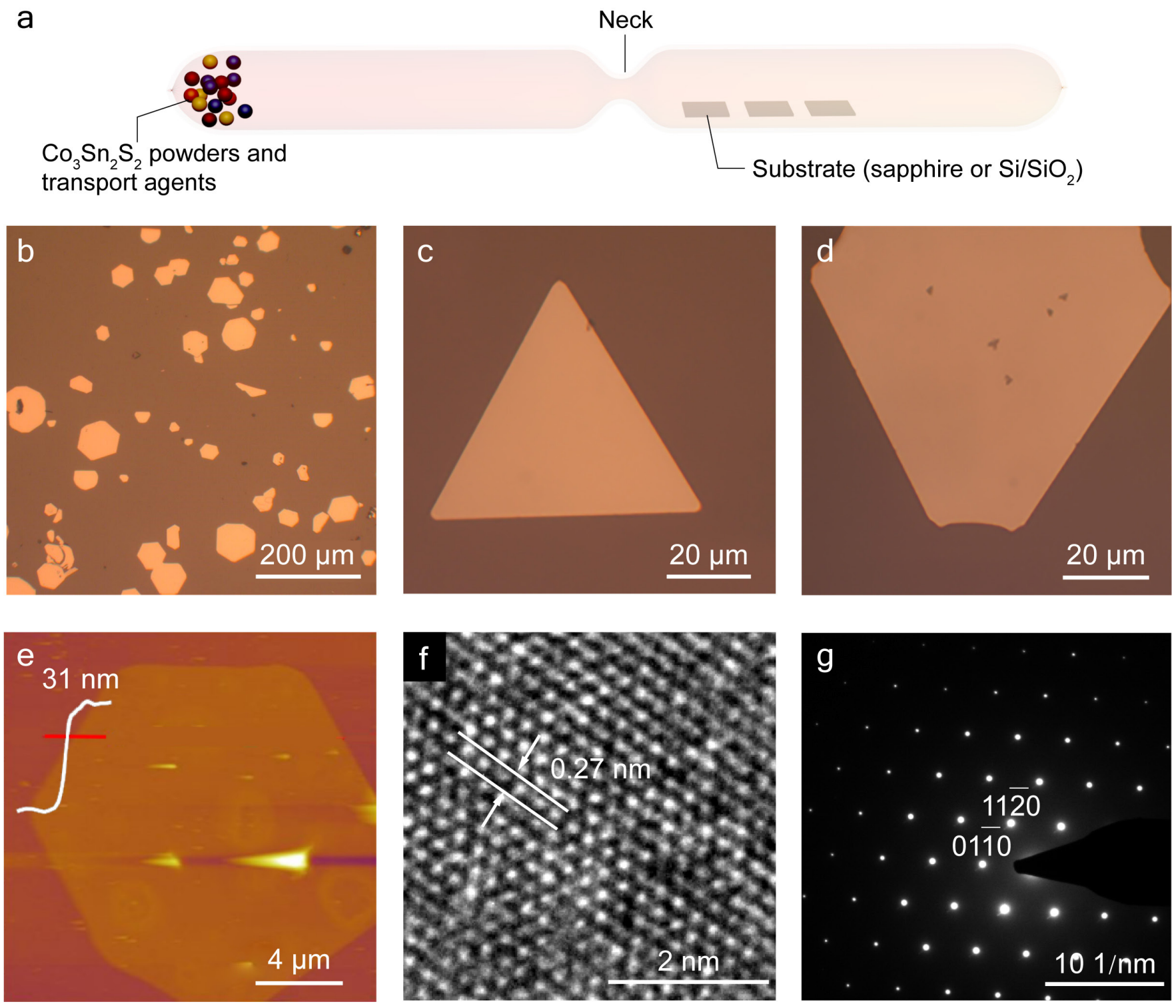

**Supplementary Fig. S1 Sample growth and crystallinity characterizations.** (a) Schematic diagram for the synthesis of $Co_3Sn_2S_2$ nanoflakes via the modified CVT method. A neck is introduced in the middle of the ampule to reduce the transport rate of the precursors. (b) Optical image of the as-grown $Co_3Sn_2S_2$ nanoflakes on a sapphire substrate. (c) Optical images of the triangle and (d) truncated hexagonal $Co_3Sn_2S_2$ nanoflakes grown on the sapphire substrates with lateral sizes up to tens of micrometers. (e) The AFM image as well as the height line profile show the thickness of $Co_3Sn_2S_2$ nanoflakes grown by this method can be as thin as 31 nm. (f) The HRTEM image of a $Co_3Sn_2S_2$ nanoflake grown on a sapphire substrate and transferred onto a lacy carbon film supported by copper TEM grids. (g) The selected area electron diffraction pattern shows a single set of the hexagonal pattern, indicating that the $Co_3Sn_2S_2$ nanoflake is a single crystal.

## Note 2: Electrical and magnetic properties

Figures S2a-b show the electric resistivity and Hall angle data measured from device 1, both are close to reported values[1,3,5]. The residual resistivity ratio (RRR, $\rho_{300K}/\rho_{2K}$) of our $Co_3Sn_2S_2$ nanoflakes ranges from 3.1 (device 1) to 12.9 (device 4). The transport spin polarization at the Fermi level is reported to be $P \approx 0.5$ at low temperature ( ~ 1 K) from point-contact Andreev reflection experiments[6,7].

The magnetic properties of $Co_3Sn_2S_2$ are listed in Table S1. Considering the small size of the nanoflake and the difficulty to transfer it, we measured the saturated magnetization of the bulk $Co_3Sn_2S_2$ sample from which the nanoflakes are grown (Fig. S2c). It is a reasonable approach since both electrical and magnetic properties such as resistivity, RRR, coercive field, magnetic anisotropy and the mechanism for reversal do not show an obvious discrepancy between the bulk and nanoflake samples, the thickness of the nanoflakes is not approaching the 2D limit, and $Co_3Sn_2S_2$ has quasi-2D character in its crystalline, electronic and magnetic properties[7].

As shown in Fig. 2d, the magnetic anisotropy field is obtained by assuming a uniaxial (*z*-axis) magnetic anisotropy in the Stoner-Wohlfarth model:

$$E = K_0 + K_u \sin^2\theta - \mu_0 \boldsymbol{H} \cdot \boldsymbol{M}$$

where $K_u$ is the uniaxial anisotropy energy, and $\theta$ is the angle between the magnetization $\boldsymbol{M}$ and the *z*-axis. The magnetic field $\boldsymbol{H}$ is applied along the *x*-axis with a 1.5-2.0° misalignment to the *z*-axis due to the leveling error from the rotation probe. The Hall resistance $R_{yx}$ as a function of magnetic field (red lines in Fig. S2d) should be proportional to the *z*-component of magnetization[10], which results to be well-fitted with results from the Stoner-Wohlfarth model (blue lines) when $K_u$ is properly chosen. Then the anisotropy field $H_K = 2K_u/\mu_0 M_s$ can be given, as well as the quality factor $Q = K_u/2\pi M_S^2$ (~200), indicating $Co_3Sn_2S_2$ a very hard ferromagnet.

To give an estimation of the domain wall width and predict the current and field-induced domain wall motion, we extract the exchange stiffness $A$ from an improved method of *M-T* curve fitting followed by a mean-field approximation[8,9]. Then, the domain wall width constant $\Delta$ can be given by $\sqrt{\frac{A}{K_u + 2\pi M_S^2}} \approx \sqrt{\frac{A}{K_u}}$, if we assume a Néel or Bloch type domain wall[10]. The extracted value is on the order of 1 nm, suggesting very thin domain walls in $Co_3Sn_2S_2$.

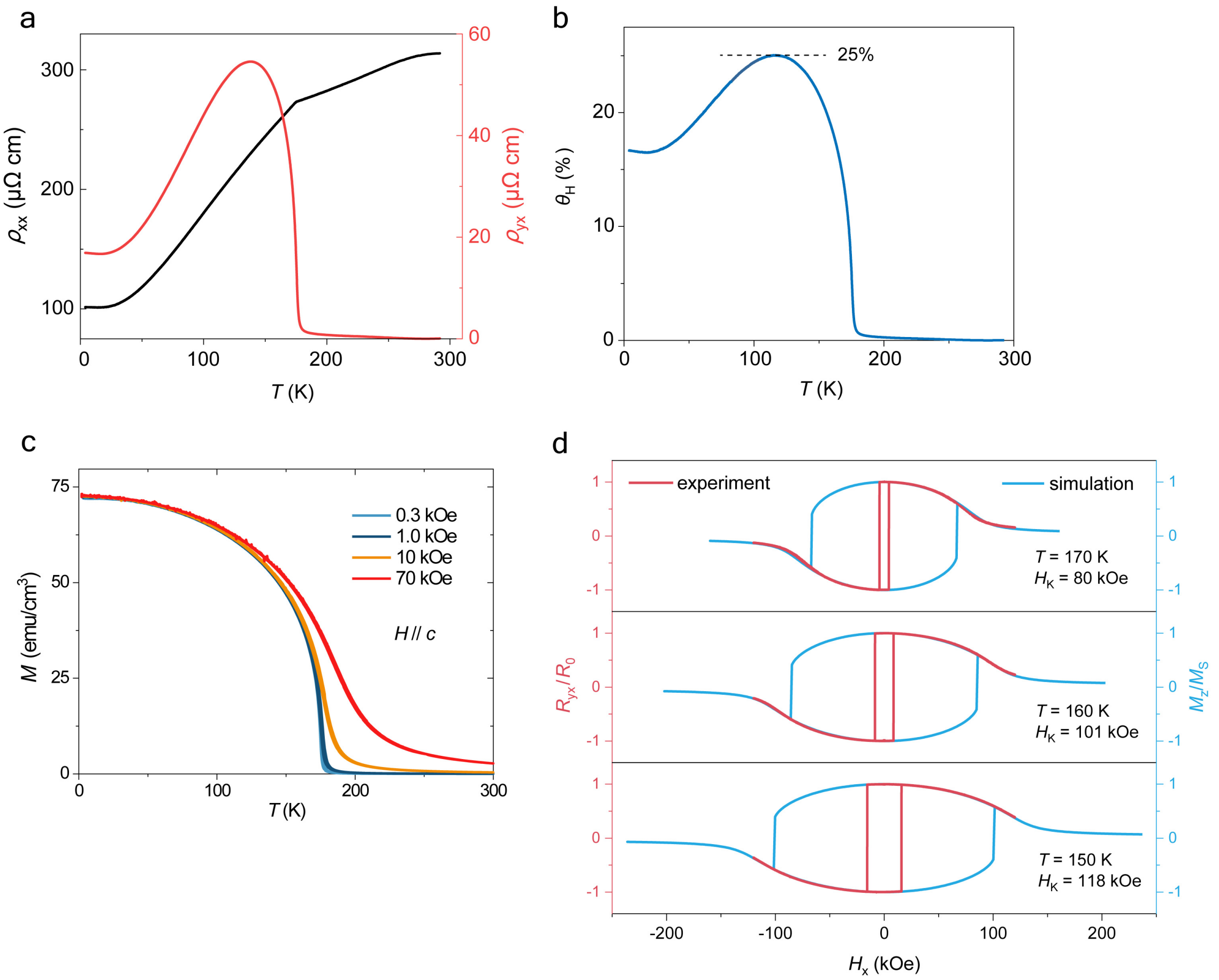

**Supplementary Fig. S1 Electrical and magnetic properties characterizations.** (a) Temperature dependence of longitudinal ($\rho_{xx}$) and transverse ($\rho_{yx}$) electric resistivity. (b) Temperature dependence of the anomalous Hall angle $\theta_H = \arctan\left(\frac{\rho_{yx}}{\rho_{xx}}\right)$ under 0.1 kOe field cooling along the *c*-axis. A peak of $\theta_H$ = 25.0% is obtained at *T* = 117 K. (c) Magnetization *M* as a function of temperature and external magnetic field. (d) Magnetic anisotropy measurement. Red curves: Hall resistance hysteresis loops with the magnetic field applied along the *x*-direction; Blue curves: calculated loops assuming a Stoner-Wohlfarth model. Because DW motion dominates the magnetization reversal process, the experimentally measured coercive field is smaller than the simulated one.

**Supplementary Table S1. Magnetic properties of $Co_3Sn_2S_2$.** $M_S$, the saturated magnetization. $H_C$, the coercive field (field sweep rate ≈ 6 Oe/s). $H_K$, the magnetic anisotropy field. $K_u$, the uniaxial anisotropy energy. *A*, the exchange stiffness. *Δ*, the magnetic domain wall width constant.

| *T* (K) | $M_S$ (emu/cm$^3$) | $H_C$ (kOe) | $H_K$ (kOe) | $K_u$ (erg/cm$^3$) | *A* (erg/cm) | *Δ* (nm) |
|---|---|---|---|---|---|---|
| 150 | 48.8 | 1.90 | 118 | $2.88\times10^6$ | $5.32\times10^{-8}$ | 1.36 |
| 160 | 41.6 | 1.29 | 101 | $2.10\times10^6$ | $3.86\times10^{-8}$ | 1.36 |
| 170 | 29.3 | 0.51 | 80 | $1.17\times10^6$ | $1.92\times10^{-8}$ | 1.28 |

## Note 3: Data on current modulation of $H_C$

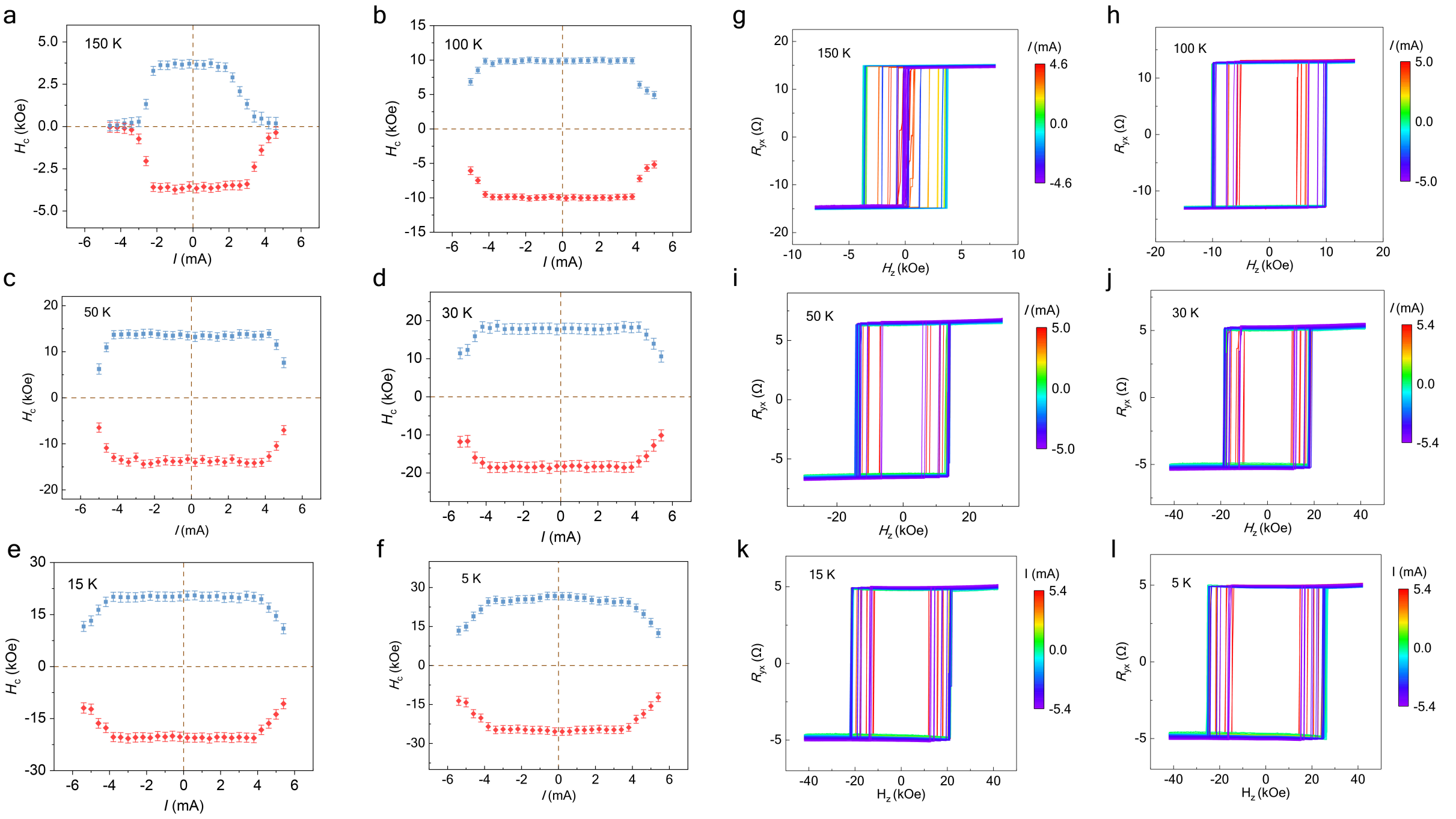

**Supplementary Fig. S2 Current modulation of coercive field and hysteresis loop at different temperatures in device 1.** (a-f) DC current dependent coercive fields $H_C$ measured at different temperatures. Current thresholds of $I_C$ = 2.1-4.4 mA were observed for all the temperatures. (g-l) DC current dependent hysteresis loops measured at different temperatures.

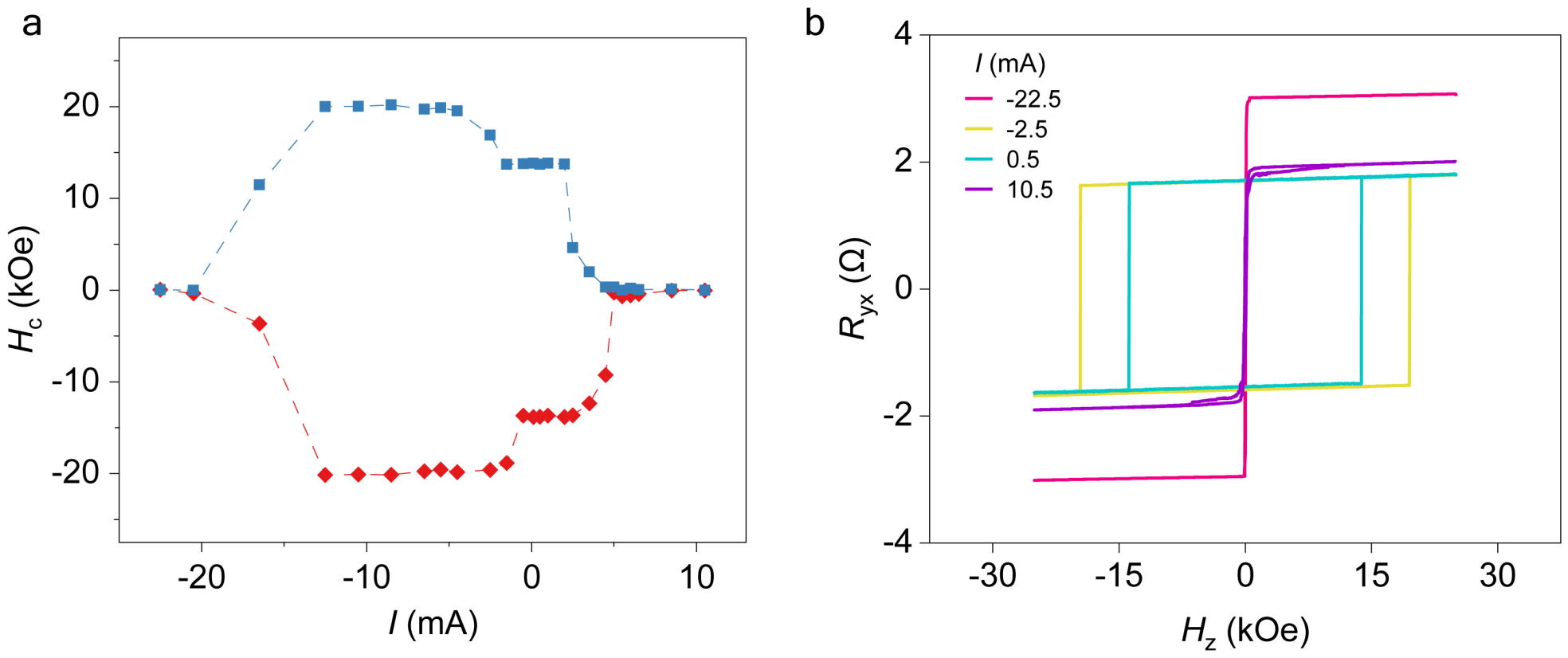

**Supplementary Fig. S3 Large $H_C$ reduction in device 2.** (a) $H_C$ as a function of current measured at 80 K. (b) Selected hysteresis loops under different DC currents. The −22.5 mA current reduces $H_C$ from 20 kOe to smaller than 0.1 kOe, so does the +10.5 mA current reducing $H_C$ of 14 kOe. The increase in $R_{yx}$ at large currents is due to Joule heating (below 40 K estimated from the $R$-$T$ curve). However, as described in the main text, the asymmetric current threshold behavior and the quantitative agreement between the efficiency of $H_C$ reduction and DW velocity measurements suggest that the giant $H_C$ reduction still mainly comes from the current-assisted DW motion.

**Note 4: Influence of thermal effects**

We evaluate the heating effect by utilizing $R_{xx}$ as an internal thermometer. Furthermore, all the temperature change is cross verified by using $R_{yx}$-$T$ curve. Here $R_{yx}$ is extracted from the half-height of the hysteresis loop at $H = 0$ so that the $R_{xx}$ component induced by the misalignment of the Hall bar electrodes is eliminated. $R_{yx}$ as a function of applied DC current $I$ is shown in Fig. S5a. Thanks to the giant anomalous Hall angle, the $R_{yx}$-$T$ curve owns a large slope (main text Fig. 1c). Therefore, by comparing $R_{xx}$ or $R_{yx}$ as a function of $I$ and $T$, we can accurately obtain the average temperature rise $\Delta T_{avg}$ as a function of $I$.

We extract $\Delta T_{avg}$ at the threshold current as a function of ambient temperature $T$ in Fig. S5b. Obviously, $\Delta T_{avg}$ is much smaller than $T_C$-$T$, so overall heating can hardly affect the magnetic reversal process. When we consider the possibility of bad contact and large boundary thermal resistance near the electrodes, severe Joule heating might happen in a small area, therefore affecting the DW nucleation process. However, in this case, $\Delta T_{avg}$ should be proportional to $T_C - T$ with a scale factor, which does not match the nonlinear $\Delta T_{avg}$-$T$ curve. More importantly, even local Joule heating does not tell the direction of the current, but most of our results show significant reliance on the DC current direction (Fig. 2 in the main text). Also, Joule heating can never result in coercive field increase as we observed in device 2 (Fig. 2d). These factors are strong evidence supporting that thermal effects are not the major mechanism in the magnetism modulation process.

By interpolating the $\Delta T_{avg}$-$I$ in the $H_C$-$T$ curve (Fig. S5c), we can estimate $\Delta H_C$ induced by overall heating effect as a function of $I$ (Fig. S5d-i). Surprisingly, at most temperatures, the $\Delta H_C$ given by $\Delta T_{avg}$ is even larger than the experiment data below the current threshold, which means an overestimation of Joule heating. This could be caused by inhomogeneous temperature rise when the original nucleation position happens to be located in the less heated area.

To eliminate the influence of thermal effect and extract the pure current effect, we plotted $\left|\Delta H_{\mathrm{C}}^{\mathrm{exp}} - \Delta H_{\mathrm{C}}^{\mathrm{thermal}}\right|$ in Fig. S6, where $\Delta H_{\mathrm{C}}^{\mathrm{thermal}}$ and $\Delta H_{\mathrm{C}}^{\mathrm{exp}}$ are the changes of the coercive field caused by the thermal effect and the experimental observation, respectively. As expected, the values are close to zero before the current threshold and start rising rapidly when the current exceeds the threshold. The current modulation efficiency $\varepsilon = \Delta H / \Delta J$ is then given by a linear fit above the current threshold (black dashed lines in Fig. S6).

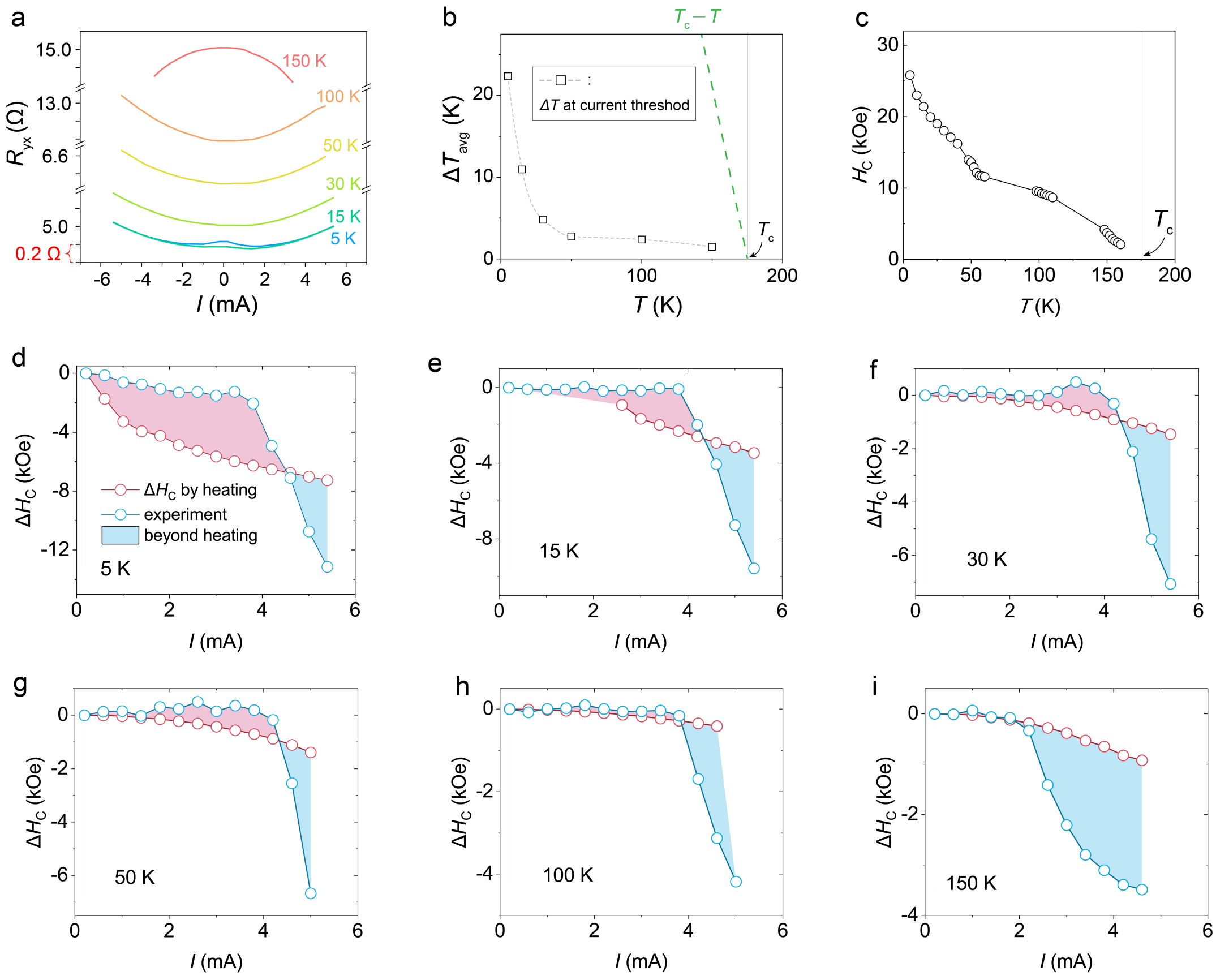

**Supplementary Fig. S5 Evaluation of $\Delta T_{avg}$ and $\Delta H_C^{thermal}$.** (a) $R_{yx}$ as a function of DC current $I$ at different temperatures. (b) Average temperature rises $\Delta T_{avg}$ at the threshold current as a function of ambient temperature $T$. (c) $H_C$ as a function of $T$. (d-i) Comparison between $\Delta H_C$ by heating and the experimental data at various temperatures. The blue shadow represents $\Delta H_C$ that cannot be explained by the thermal effect.

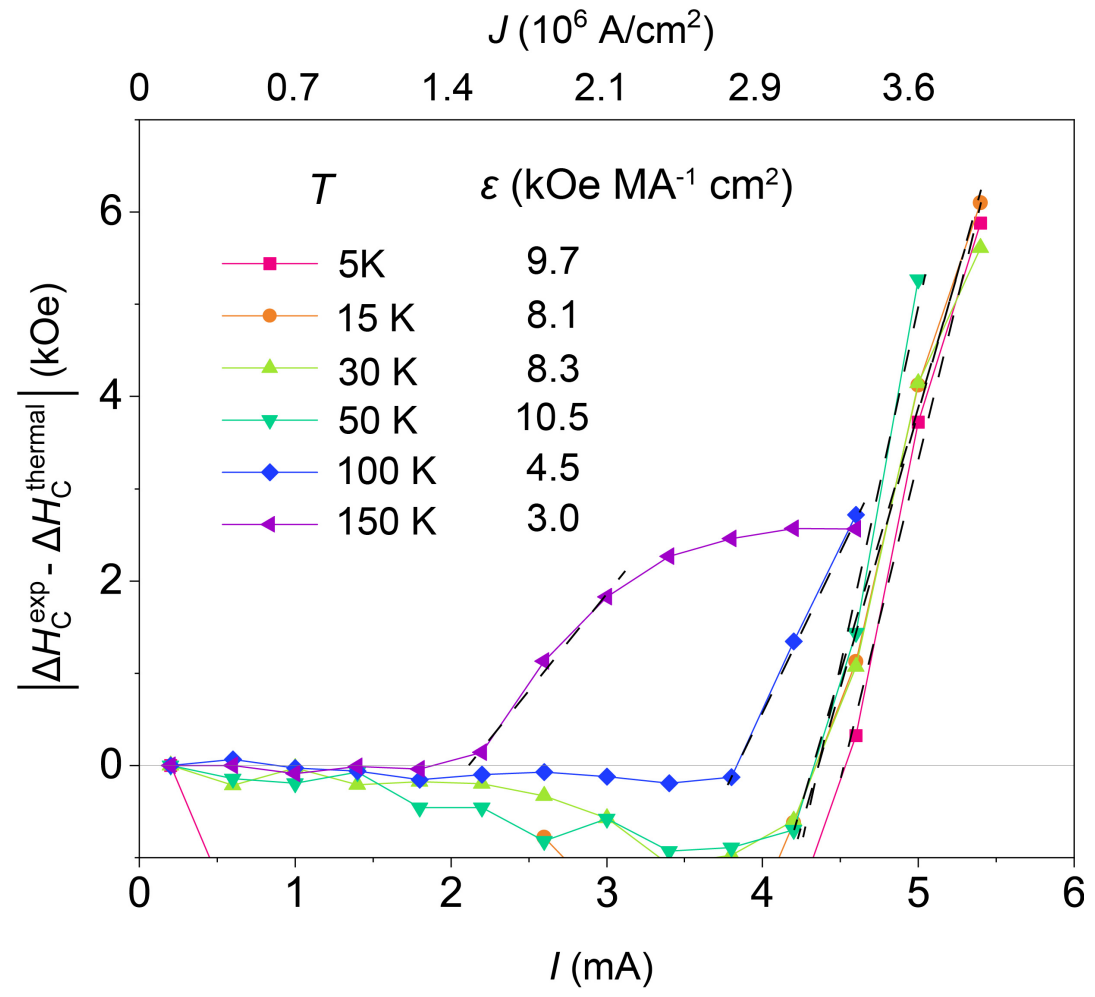

**Supplementary Fig. S6 Extracted $\Delta H_C$ beyond thermal effect.** Values of $\left|\Delta H_C^{exp} - \Delta H_C^{thermal}\right|$ as a function of current (density) $I$ ($J$) are calculated from the data in Fig. 5(d-i).

The black dashed lines represent linear fits above the current threshold, which correspond to the authentic current modulation efficiency $\varepsilon=\Delta H/\Delta J$.

**Note 5: Symmetry for preparing a magnetic domain wall**

In the DW velocity measurement, a DW is generated by the mutual effect of the current pulse on the heater, the external magnetic field, and the source-drain current. Here we elucidate this process by showing how these parameters affect the nucleation result.

First, we measured the $I_{sd}$-$H_z$ phase boundary for magnetization reversal. The experimental setup is shown in Fig. S7a. Under the same nucleation process as in the DW velocity measurement, we monitored Hall signals at three distinct Hall bars simultaneously after the current pulse is applied on the heater. The results are shown in Fig. S7b-d. Since the sample is initially up-magnetized, the blue region in the figure represents that the magnetization at the Hall bar stays unchanged, while the red region indicates full local magnetization reversal detected by the Hall resistance. Basically, the nucleation field will decrease as the current towards the heater increases, as expected (more detailed discussion in Note 8). In addition, we observed that:

- Three Hall bars give the same phase boundary, indicating that the DW will not get stuck in the nanowire after being freed near the heater, as well as that the nanowire does not have additional nucleation sites between Hall bars.
- At some points, the magnetization is partly reversed at Hall bar 1, close to the phase boundary (the white region in Fig. S7b), while the magnetizations at Hall bar 2 and 3 are not reversed. Because Hall bar 1 is next to the heater, a possible reason is the thermal instability of domain walls close to the critical nucleation condition. Domain walls could be broken or tilted, therefore cannot continue propagating.
- A finite negative magnetic field is essential for DW nucleation. One possibility is that the current pulse on the heater is too small to produce an Oersted field, so a finite field is required to disturb the magnetization and enable the source-drain current to generate STT that helps the DW nucleation (see Note 8). Another possible reason is that the source-drain current only exerts a force on the DW in the $x$-direction (length), so the DW expansion in the $y$-direction (width) must be fulfilled by a magnetic field.
- The nucleation field saturates at about −1.4 kOe for large negative currents, equal to the coercive field without heater and $I_{sd}$. This is because DW nucleation happens at other positions in the sample, after which the whole magnetization is reversed.
- The nucleation phase boundary does not correspond to a threshold for DW motion. Indeed, the linear dependence of $v_{DW}$ on $I_{sd}$ and $H_z$ keeps well down to the lowest $I_{sd}$ and $H_z$ value we have tested, and the extrapolation of this linear dependency gives $v_{DW} = 0$ at $(I_{sd}, H_z) \approx 0$ (see Fig. S10a,b), but not at the nucleation phase boundary that we have observed here.

Second, we gave a full description of how the directions of the parameters affect DW nucleation, from which we can distinguish different current effects including Joule heating, current-induced Oersted field, SOT, and STT. The nucleation results are shown in Fig. S8 and Table S2. In summary, in order to nucleate the DW at moderate $I_{sd}$ and small $H_z$ (much smaller than the intrinsic coercive field), we need:

1. The external field $H$ to be in opposite direction with the initial magnetization $M$.
2. The source-drain current $I_{sd}$ to flow towards the heater.

If and only if the parameters obey the two rules, a DW can be successfully nucleated after $I_{heater}$ is applied. Other configurations, including whether $M$ is initially up or down, the direction of $I_{heater}$, and the location of the heater do not affect the conclusion. From these observations, we are able to analyze different current effects on the nucleation process as follows.

For $I_{heater}$, only the heating effect is important because under all circumstances its direction will not change the nucleation result, nor the $I_{sd}$-$H_z$ map.

The Oersted field generated by $I_{sd}$ could be ignored, because, for nucleation, $I_{sd}$ always flows in the same direction (towards the heater), independent of the initial $M$ direction. Meanwhile, the magnitude of the potential Oersted field is much smaller than the change of $H_z$, under our sample geometry and current density[11].

The possible SOT generated by $I_{sd}$ is also negligible. We illustrate this by several points:

1. In a Weyl semimetal where strong SOC is expected in its surface states (Fermi arcs) and bulk bands near the Fermi surface[12], the application of an electric field could induce spin polarization at the surface and in the bulk, and therefore generate SOT on the local magnetization[13,14]. The crystal structure of $Co_3Sn_2S_2$ (space group *R-3m*) is invariant under inversion, so in general, the SOT from bulk states is prohibited[15,16]. For top and bottom surfaces, because of the existence of the substrate ($SiO_2$ or sapphire), the inversion symmetry could be broken, and non-zero SOT is possible[17–19]. However, in our experimental configuration where the external magnetic field is applied along *z*-direction and the current is in the *x*-direction, the effective field generated by the field-like SOT $\boldsymbol{\tau}_{\mathrm{FL}} \propto \boldsymbol{M} \times \hat{\boldsymbol{y}}$ or the damping-like SOT $\boldsymbol{\tau}_{\mathrm{DL}} \propto \boldsymbol{M} \times (\hat{\boldsymbol{y}} \times \boldsymbol{M})$ is an even function of $I_{sd}$, so the nucleation result should not depend on the direction of $I_{sd}$. Also, if SOT dominates, the $H_z$-$I_{sd}$ phase boundary will be symmetric about $I_{sd}$, and $H_z$ should monotonically decrease as the absolute value of $I_{sd}$ increases, which is fundamentally different from the experimental results.
2. The effective thickness of surface (2D) conducting channels in $Co_3Sn_2S_2$ is reported[20] to be about 20 nm. On contrary, the current modulation behavior is consistently observed in all tested samples with various thicknesses ranging from 35-165 nm. In addition, the current efficiencies of these samples are independent of the thickness (main text Fig. 4), implying the 3D nature of the current effect.
3. The SOT comes from the intrinsic band structure, so it should not depend on the position of the heater.

These contradictions exclude the possibility that SOT dominants our experimental results.

**Supplementary Table S2. Symmetries required for DW nucleation (further to Fig. S8).** The marks of directions (+ or -) are the same as shown in Fig. S8. The +/- and L/R notations indicate that either option gives the same nucleation result.

| $M$ | $H$ | $I_{sd}$ | $I_{heater}$ | Heater position | Nucleation |
|---|---|---|---|---|---|
| - | + | + | +/- | R | Yes |
| + | - | + | +/- | R | Yes |
| - | + | - | +/- | R | No |
| + | - | - | +/- | R | No |
| - | + | + | +/- | L | No |
| + | - | + | +/- | L | No |
| - | + | - | +/- | L | Yes |
| + | - | - | +/- | L | Yes |
| - | - | +/- | +/- | L/R | No |
| + | + | +/- | +/- | L/R | No |

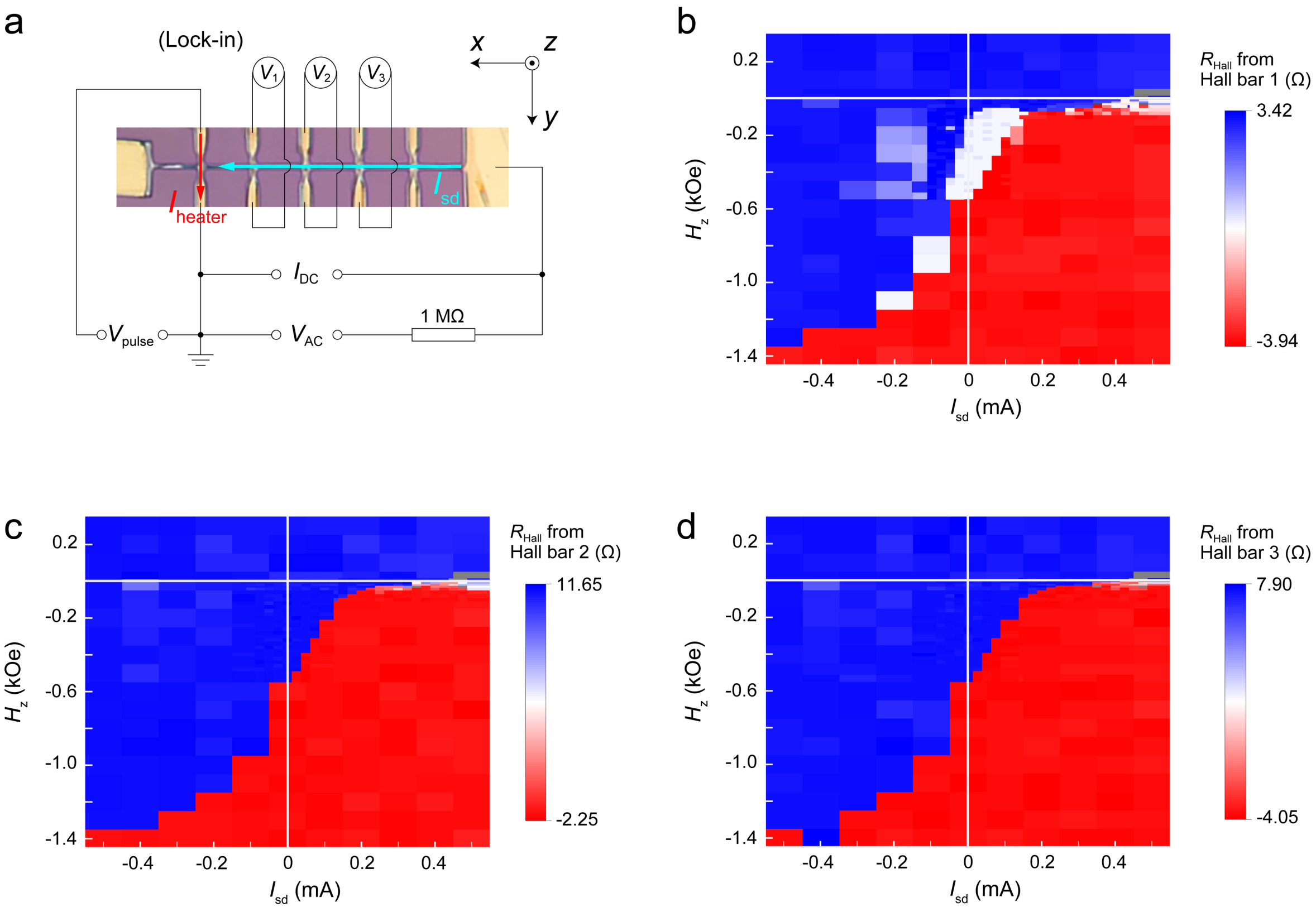

**Supplementary Fig. S7 $I_{sd}$-$H_z$ phase boundary for DW nucleation.** (a) Measurement setup. The Hall voltages at three distinct Hall bars are monitored by Lock-in amplifiers. (b-d) $I_{sd}$-$H_z$ phase boundaries of magnetization reversal at Hall bars 1, 2, and 3. Each pixel represents an $R_{Hall}$ value obtained under $I_{sd}$ and $H_z$, after applying a current pulse $I_{heater}$ on the left heater. The positive $I_{sd}$ direction is from right to left (towards the heater), as shown in (a). The red and blue regions represent magnetization reversed and not-reversed, respectively.

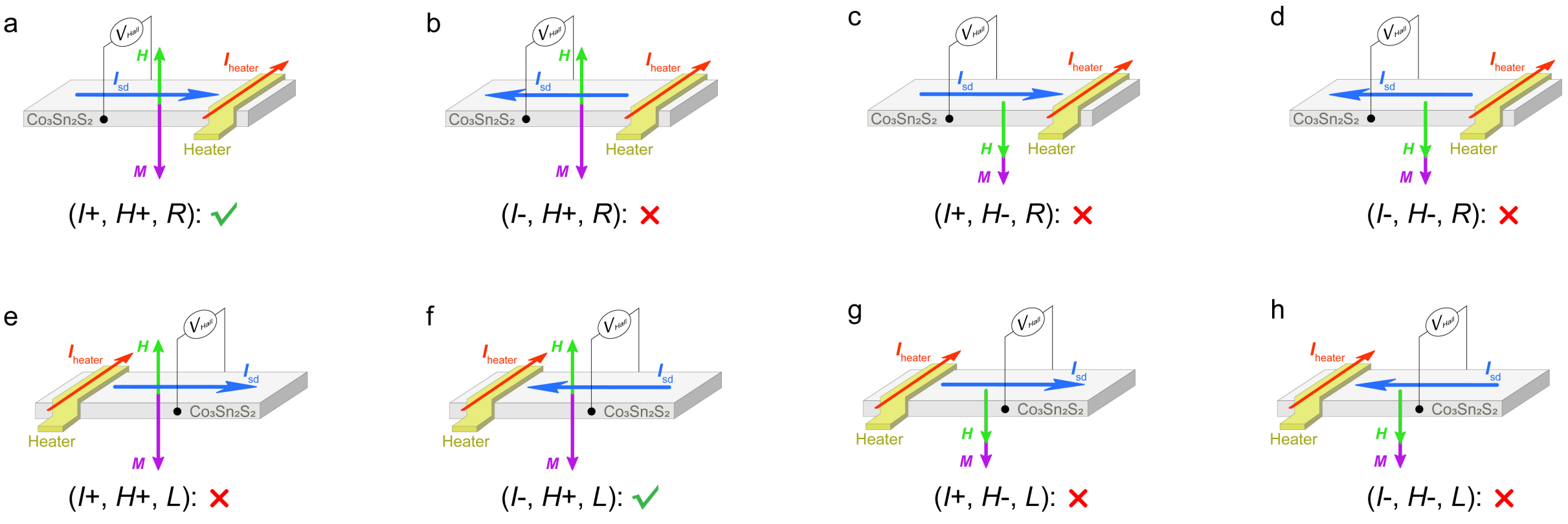

**Supplementary Fig. S8 Symmetries required for DW nucleation.** The figures show the directions of source-drain current $I_{sd}$, external field $H$, initial magnetization $M$, and the heater position relative to the Hall probe ($L$ for left and $R$ for right) that are required for successful DW nucleation, under small $H$ and moderate $I_{sd}$, e.g., $|I_{sd}|$=0.3 mA and $|H|$= 0.2 kOe in Fig. S7. The direction of $I_{heater}$ does not affect, and the initial magnetization only reverses the required eternal field direction. A full comparison is shown in Table S2.

## Note 6: Discussion on domain wall velocity and mobility

The domain wall velocity data of device 5 is presented in Fig. S9d-f. For comparison, the data of device 4 (used in the main text) is shown again in Fig. S9a-c. The results are rather close. The linear dependency of $v_{DW}$ on $H_z$ and $J$ claimed in the main text holds on device 5. The current- and field-dependent DW mobility $\eta$ and $\mu$ of the two samples are identical within a factor of 2. Also, the trending of their $\eta$-$T$ and $\mu$-$T$ curves are qualitatively consistent, where $\mu$ increases as $T$ increases, while $\eta$ is less sensitive to temperature change.

Next, we checked the validity of the formula $v_{DW} = \eta J + \mu H_z$ by fitting the experimental data of $v_{DW}(J, H_z)$ to $v_{DW} = \eta(J - J_c^{fit}) + \mu H_z$, where $J_c^{fit}$ is the fitted threshold current density. The results are as follows:

| Sample | $\eta$ / $10^{-9}$ m$^3$C$^{-1}$ | $\mu$ / $10^{-3}$ m s$^{-1}$Oe$^{-1}$ | $J_c^{fit}$ / $10^5$ A/cm$^2$ | $R^2$ (COD) |
|---|---|---|---|---|
| device 4 | 0.349 ± 0.003 | 1.750 ± 0.018 | 1.09 ± 0.05 | 0.987 |
| device 5 | 0.491 ± 0.007 | 1.082 ± 0.062 | –0.55 ± 0.12 | 0.981 |

Both samples fit the linear function well. The experiment data and the fit planes are shown in Fig. S10a-b. Interestingly, data from device 5 gives a negative $J_c^{fit} = (-0.55 \pm 0.12)\times10^5$ A/cm$^2$. This negative value is supposed to come from the suppression of $v_{DW}$ (therefore deviates from the linear function) at higher velocities[21], because we measured $v_{DW}$ of device 5 at moderate $H_z \geq 0.2$ kOe. Taken into consideration that the $v_{DW}(J)$ curve around the threshold current cannot be described by the linear relationship (which is only applicable in the flow region), these results only serve as a rough lower-bound estimation of the field-free threshold current density in $Co_3Sn_2S_2$. An upper bound of $J_c$ is provided by the stochastic DW nucleation at $H_z = 0$ kOe and $J = 5.1\times10^5$ A/cm$^2$ followed by DW propagation with $v_{DW}>0$ observed in device 5 (Fig. S9f). Remind the effective threshold current density $J_c^*$ deduced in the main text under moderate magnetic field, we now have three estimations of the threshold current density:

- A lower bound of $J_c = J_c^{fit}$ ranging from (0-1.09)×$10^5$ A/cm$^2$ derived from linear fit and extrapolation.
- An effective $J_c = J_c^*$ ranging from (1.5-2.8)×$10^5$ A/cm$^2$ given by $J + (\mu/\eta) H_z$ at moderate $H_z = 0.2$ kOe.
- An upper bound of $J_c < 5.1\times10^5$ A/cm$^2$ was observed directly at $H_z = 0$ kOe.

Because the latter two thresholds were given at quite large $v_{DW} \sim$ 1-3 m/s, they do not prevent a lower intrinsic threshold current density. In a perfect magnetic Weyl semimetal, the intrinsic $J_c$ is predicted to vanish (see Note 1). Our data could not exclude this possibility, which we believe is worthy of further experimental verification.

In addition to the fundamental physics interest, the magnetic and DW motion parameters of $Co_3Sn_2S_2$ are also highly desirable in the design of functional spintronics devices. The current-dependent DW mobility $\eta = \Delta v_{DW}/\Delta J$ reaches $0.5\times10^{-9}$ m$^3$/C, close to the highest value (~ $0.71\times10^{-9}$ m$^3$/C) reported in STT-driven systems[22]. The low field-dependent DW mobility $\mu < 2.3\times10^{-3}$ m/(s·Oe) strengthens the device reliability under magnetic field exposure. The ultra-high STT efficiency $\Delta H/\Delta J = \eta/\mu$ (larger than 10 kOe MA$^{-1}$ cm$^2$ at 80 K) further enables magnetic-field resistive and high-performance DW devices.

We extended the DW velocity data to negative current and both up and down initial magnetization direction. The results are shown in Fig. S11. The negative velocity at a positive current indicates that the DW moves in the electron flow direction. The up-to-down and down-to-up DWs move in the same direction, and the velocity direction reverses at a negative current, demonstrating that the DW is STT driven.

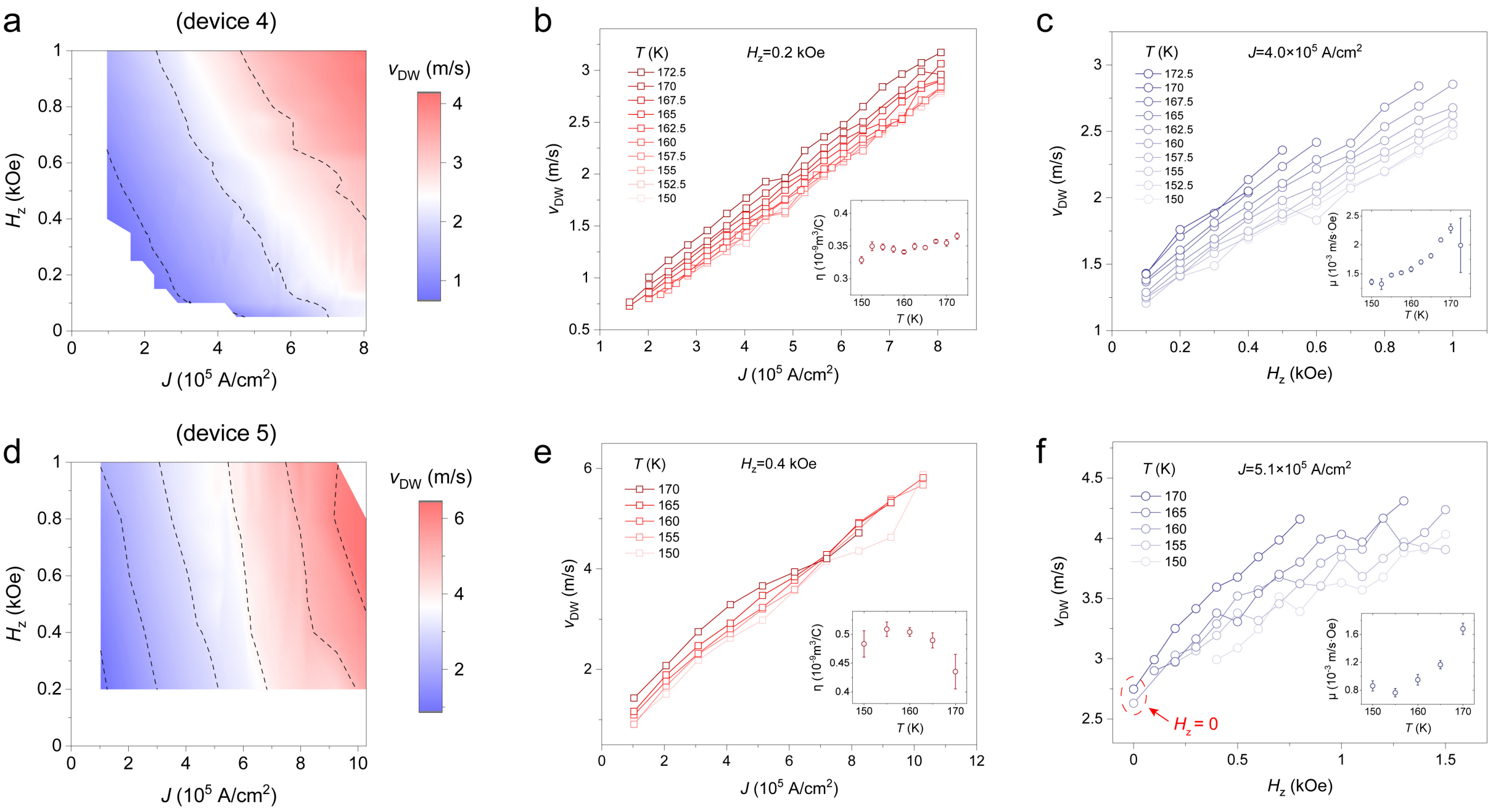

**Supplementary Fig. S9 Comparison of DW velocity data between two devices.** (a-c) Data from device 4, which is the one presented in the main text (90 μm×1 μm×62 nm). (d-f) Data from another $Co_3Sn_2S_2$ nanowire (device 5, 20 μm×1.2 μm×81 nm). (a,d) Contour map of $v_{DW}$ as a function $H_z$ and $J$. (b,e) $v_{DW}$ as a function of $J$ and $T$, under the fixed value of $H_z$. Inset, the current-dependent DW mobility $\eta$ as a function of $T$. (c,f) $v_{DW}$ as a function of $H_z$ and $T$, under the fixed value of $J$. Inset, the field-dependent DW mobility $\mu$ as a function of $T$.

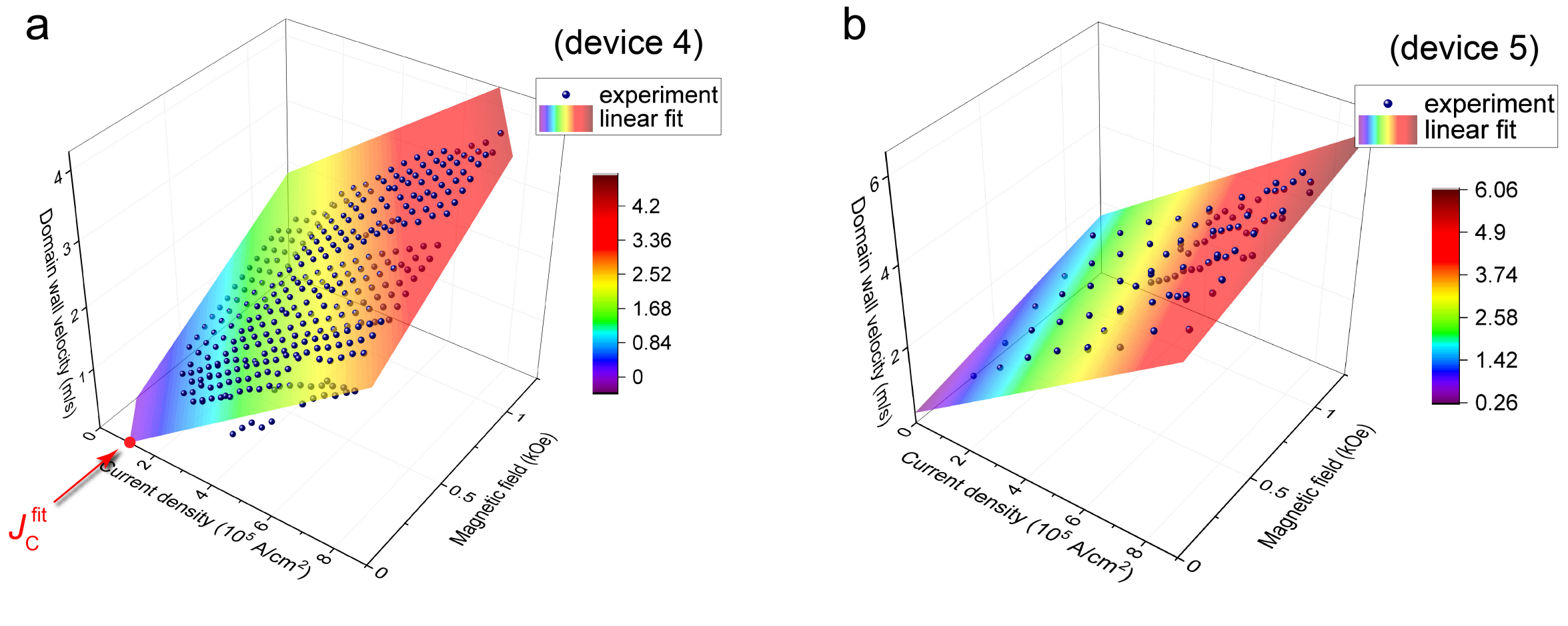

**Supplementary Fig. S10** (a,b) The linear fit result of $v_{DW} = \eta(J - J_c^{fit}) + \mu H_z$. The blue dots are experimental data, and the colored plane is the fitted surface. For device 4, $J_c^{fit}$ is positive, which is marked in (a). Device 5, on the contrary, has a negative $J_c^{fit}$ value.

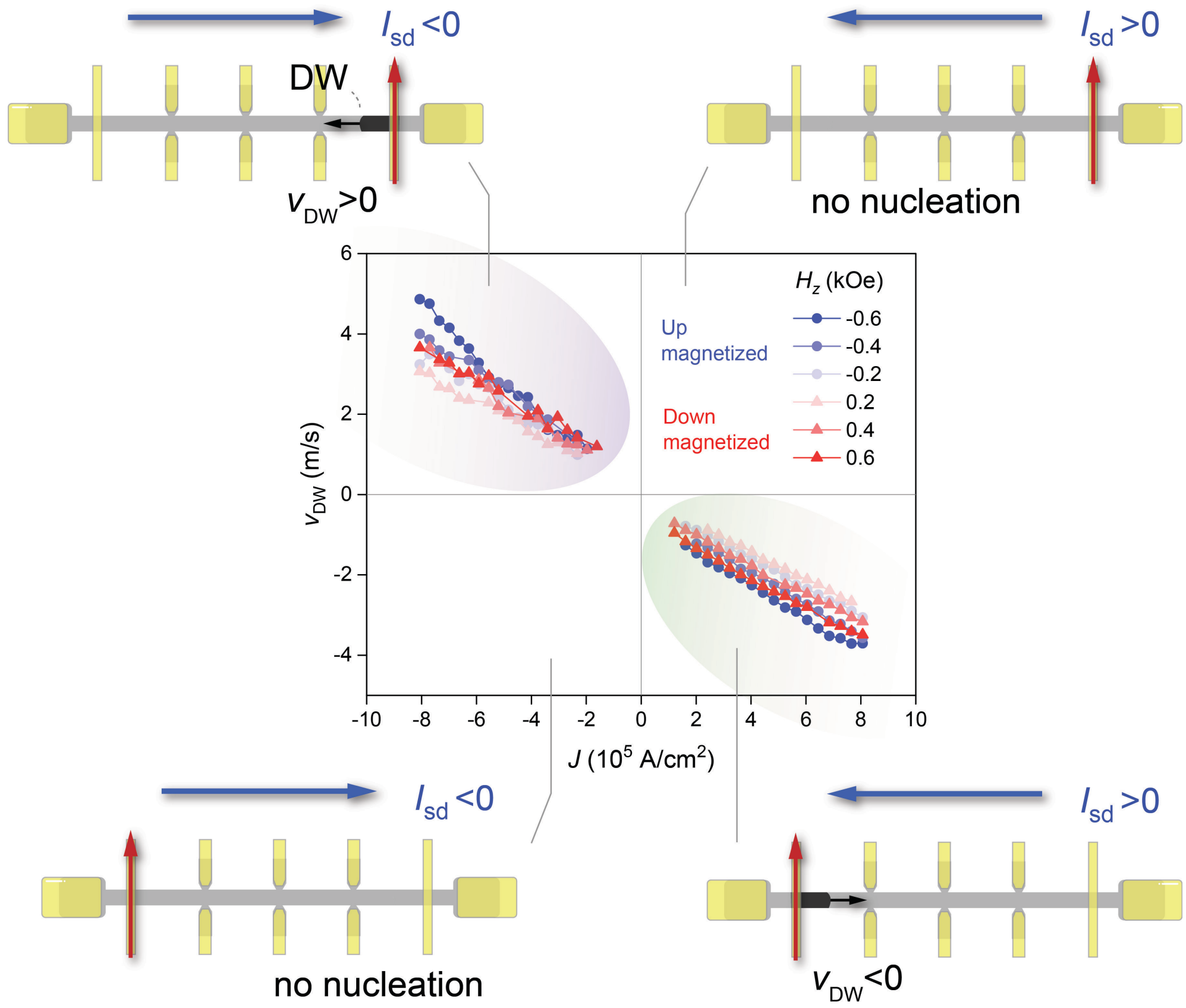

**Supplementary Fig. S11** $v_{DW}$ as a function of $J$, $H_z$, and initial magnetization direction (device 4 at 160 K). The blue and red data points represent down-to-up and up-to-down domain walls, respectively.

## Note 7: Discussion on the equivalence of the two efficiencies

In the main text, we confirmed the current-assisted DW motion model by comparing the experimental value of $H_C$ modulation efficiency and the STT effective field on a mobile domain wall. Here, we provide further discussion on the equivalence of the two efficiencies.

As stated earlier, $H_C$ equals the field that can "free" a DW from a particular nucleation site, which is usually at corners or near imperfections of a ferromagnet[23]. This process can be categorized into two situations depending on whether the DW exists or not before it is freed.

In the first case, small domains of inversed magnetization are already present near-surface bumps or sharp corners, to reduce the density of free poles and thus lower the magnetostatic energy. As a result, $H_C$ is determined by the DW depinning process, which can be described by the Arrhenius law $\tau = \tau_0 \exp\left(\frac{E_b}{k_B T}\right)$, where $E_b$ is the energy barrier that pins the DW, $\tau$ is the characteristic time for depinning, $1/\tau_0$ is the attempt frequency, $T$ is the temperature, and $k_B$ is the Boltzmann constant. In the case without current, the energy barrier can be written as $E_b(H_z) = E_{b,0}\left(1 - \frac{H_z}{H_{dep}}\right)$, where $H_{dep}$ is the depinning field.

From the DW velocity measurement, the current linearly acts as an effective magnetic field on the domain wall. It can also be seen in the LLG equation for a 1D DW moving in the $x$-direction: $\frac{d\boldsymbol{M}}{dt} = -\gamma \boldsymbol{M} \times \boldsymbol{H}_{\text{eff}} + \frac{\alpha}{M_S}\boldsymbol{M} \times \frac{d\boldsymbol{M}}{dt} - u\frac{\partial \boldsymbol{M}}{\partial x} + \frac{\beta u}{M_S}\boldsymbol{M} \times \frac{\partial \boldsymbol{M}}{\partial x}$, where the last two terms represent the adiabatic and non-adiabatic STT, $u = \frac{gJ\mu_B P}{2eM_S}$, and $\beta$ is the non-adiabatic STT parameter. The $\frac{\beta u}{\gamma M_S}\frac{\partial \boldsymbol{M}}{\partial x}$ part in the non-adiabatic STT term acts just as a magnetic field, which is predicted to dominate in the Weyl model. From these observations, the energy barrier as a function of field and current density has the form[24,25] $E_b(H_z, J) = E_{b,0}\left[1 - (H_z + \varepsilon J)/H_{dep}\right]$, where $\varepsilon$ is the current efficiency. We attribute the experimentally observed current threshold to the requirement that the nucleation position has to be located near the electrodes, as shown in the control experiment in the main text Fig. 3c-h. Because of the small saturated magnetization and thus the small demagnetization field, the magnetization in the DW should enter processional motion as long as the DW is freed. So, the current efficiency $\varepsilon$ here should be the same as the efficiency when it pushes a mobile DW under the same magnitude of the field.

In the second case, the DW is generated by magnetization rotation in the nucleation area. The fact that $H_C$ is much smaller than $H_K$ in $Co_3Sn_2S_2$ implicates an anisotropy gradient at corners or near imperfections, which will lead to non-uniform magnetization in this area. As a result, the electric current still applies STT on the local magnetization. This mechanism has been reported in several papers[26,27].

Like the depinning case, the reversal process of local magnetization can also be described by the Arrhenius law $\tau = \tau_0 \exp\left(\frac{E_b}{k_B T}\right)$. Given the damping-like STT in a spin valve structure

$$\boldsymbol{T} \propto J\boldsymbol{M} \times \left(\boldsymbol{M} \times \widehat{\boldsymbol{M}}_p\right) \tag{1}$$

where $J$ is the current density and $\widehat{\boldsymbol{M}}_p$ is the magnetization direction in the pinned layer, theoretical work has proved that the effective energy barrier is[28]

$$E_{\mathrm{b}}(H_z,J) \approx E_{b,0}\left(1-\frac{H_z}{H_{n,0}}\right)^s\left(1-\frac{H_z+\varepsilon' J}{H_{n,0}}\right) \quad (2)$$

where the parameter $s$=0.5-1.0, and $H_{\mathrm{n},0}$ is the nucleation field without current. In our case of inhomogeneous magnetization and in-plane current injection, eq. (2) corresponds to the adiabatic STT term. If the field-like non-adiabatic STT term dominates, the parameter $s$ will become closer to 0, thereby eliminating the non-linearity on the current-induced effective field, and we return to the same expression as depinning: $E_{\mathrm{b}}(H_z,J) = E_{\mathrm{b},0}\left[1-(H_z+\varepsilon' J)/H_{\mathrm{n},0}\right]$.

However, the current efficiency $\varepsilon'$ here is not necessarily the same as $\varepsilon$ in the depinning case. Actually, it is a function of magnetization in the nucleation area, $\varepsilon' = \varepsilon'(\langle \boldsymbol{M} \rangle)$, because the $\frac{\partial \boldsymbol{M}}{\partial x}$ term depends on the profile of inhomogeneous magnetic texture, which is continuously changing with $\boldsymbol{M}$. Under the high barrier condition $E_{b,0} \gg k_B T$, the accumulated reversal probability $F(t) = 1 - \exp\left\{-\frac{t}{\tau_0} \times \exp\left[\frac{E_b\left(H_z,J,\varepsilon'(P)\right)}{k_B T}\right]\right\}$ becomes no longer well-defined. We cannot give an explicit expression of $H_z$ required for nucleation (defined as $F(t_0) = 50\%$, where $t_0$ is the measurement time) as a function of $J$ in this case, but the current efficiency must be considerably different from the value obtained in depinning or current-driven DW motion. On the other hand, if $E_{b,0}$ is not much larger than $k_{\mathrm{B}}T$, the current efficiency can be approximately given by $\Delta H/\Delta J \approx \varepsilon'(F=0.5)$. In this case, $\varepsilon'$ can be close to the value obtained in the DW velocity measurements.

The discussion above provides an intuitive physical picture of the efficiency of coercive field modulation induced by current injection. Considering the lack of direct observation of the nucleation process (the time resolution of conventional magnetic imaging equipment is difficult to reach the order of microseconds at a sample size of several tens of micron), the complexity of the real magnetization profiles, and the considerable error in magnetic parameters such as domain wall width and exchange stiffness, we cannot comment seriously on the definitive microscopic nucleation mechanism. Nevertheless, we clarified that the STT can significantly modify the DW nucleation process and therefore, alter the nucleation field with efficiency comparable to the value in the case of current-driven DW motion.

**Note 8: Quantitative understanding of the high STT efficiency**

Here we provide a quantitative current- and field-driven DW motion model to explain the observed high STT efficiency and low current threshold in $Co_3Sn_2S_2$. As a starting point, we reinvestigate how the material's magnetic parameters will affect current- and field-dependent DW mobility. We notice recent theoretical works predict a large non-adiabatic STT in magnetic Weyl semimetals[29,30]. The influence of this Weyl fermions-induced torque is shrunk into the non-adiabatic STT coefficient $\beta$, and will be discussed hereafter.

**Supplementary Table S3.** Current-dependent DW mobility $\eta$ and field-dependent DW mobility $\mu$ of the two measured nanowire devices.

| | $T$ (K) | $\eta$ ($10^{-9}$ m$^3$/C) | $\mu$ ($10^{-3}$ m/(s·Oe)) |
|---|---|---|---|
| Device 4 | 150 | 0.328 | 1.357 |
| | 160 | 0.341 | 1.577 |
| | 170 | 0.355 | 2.283 |
| Device 5 | 150 | 0.483 | 0.861 |
| | 160 | 0.504 | 0.949 |
| | 170 | 0.435 | 1.680 |

**Supplementary Table S4.** Known magnetic parameters of $Co_3Sn_2S_2$.

| $T$ (K) | $M_S$ (emu/cm$^3$) | $K_u$ (erg/cm$^3$) | $A$ (erg/cm) | $\Delta$ (nm) |
|---|---|---|---|---|
| 150 | 48.8 | $2.88\times10^6$ | $5.32\times10^{-8}$ | 1.36 |
| 160 | 41.6 | $2.10\times10^6$ | $3.86\times10^{-8}$ | 1.36 |
| 170 | 29.3 | $1.17\times10^6$ | $1.92\times10^{-8}$ | 1.28 |

The known magnetic parameters of $Co_3Sn_2S_2$ are summarized in Table S4. However, to date the Gilbert damping factor $\alpha$ of $Co_3Sn_2S_2$ has not been reported, which can be understood because extremely high frequency field of $\gamma H_K/2\pi \sim 280$ GHz is required to perform ferromagnetic resonance experiments, greatly outside the experimentally accessible range. We could nevertheless use the DW mobility data to derive the $\alpha$ value, and therefore analyze the DW dynamics in greater detail.

Depend on the domain wall motion mode (steady or processional) in the flow region, the domain wall mobilities have different expressions[10,31]:

$$\eta = \frac{\beta}{\alpha}u,\ \mu = \frac{\gamma\Delta}{\alpha}, \qquad \text{when } \left|H + (\beta-\alpha)\frac{u}{\gamma\Delta}\right| \leq 2\pi\alpha M_S\left|N_x - N_y\right| \qquad (3.\text{a})$$

$$\eta = \frac{\alpha\beta+1}{1+\alpha^2}u,\ \mu = \frac{\alpha}{1+\alpha^2}\gamma\Delta, \quad \text{when } \left|H + (\beta-\alpha)\frac{u}{\gamma\Delta}\right| \gg 2\pi\alpha M_S\left|N_x - N_y\right| \qquad (3.\text{b})$$

Here $u = \frac{g\mu_B PJ}{2eM_S}$, $\Delta = \sqrt{A/K_u}$, $N_x$ and $N_y$ are the demagnetization factors along $x$- (current) and $y$- (width) direction. Eq (3.a) stands for steady motion of the magnetization inside the DW under small field and current density (before Walker breakdown), and Eq (3.b) represents processional motion after Walker breakdown under large field or current density. The Walker field $2\pi\alpha M_S|N_x - N_y|$ is only ~ 3 Oe, if we plug in $M_S$ = 48.8 emu/cm$^3$, $N_x$ = 0, $N_y \approx 0.1$ for the nanowire geometry, and a typical value of $\alpha = 0.1$. For comparison, the field we applied in the DW velocity measurement is 100-1000 Oe, much larger than the Walker field. We double-checked that Eq (3.b) should be used to calculate the DW mobilities by comparing the obtained $\alpha$ values using the two field-dependent DW mobility expressions (Table S5). Clearly, only Eq (3.b) gives reasonable values, so the magnetization inside the DW enters processional motion, which is also confirmed by micromagnetic simulation.

**Supplementary Table S5.** Gilbert damping factor $\alpha$ calculated by Eq (1.a) and Eq (1.b).

| | $T$ (K) | $\alpha$, assume $\mu = \gamma\Delta/\alpha$ (1.a) | $\alpha$, assume $\mu = \gamma\Delta\alpha/(1+\alpha^2)$ (1.b) |
|---|---|---|---|
| Device 4 | 150 | 17.6 | 0.057 |
| | 160 | 15.2 | 0.066 |
| | 170 | 9.9 | 0.102 |
| Device 5 | 150 | 27.8 | 0.036 |
| | 160 | 25.2 | 0.040 |
| | 170 | 13.4 | 0.075 |

Next, we can study the current-dependent DW mobility

$$\eta = \frac{\alpha\beta + 1}{1 + \alpha^2} u = \left(\frac{\alpha\beta + 1}{1 + \alpha^2}\right) \frac{g\mu_B PJ}{2eM_S} \tag{4}$$

Unfortunately, we cannot directly obtain the non-adiabatic STT coefficient $\beta$ (which is most closely related to the "Weyl physics") using $\eta$, because the transport spin polarization $P$ is another unknown parameter. The reported value of $P \approx 0.5$ at $T \sim 1$ K is not accurate enough to reveal any solid information about $\beta$. The spin mixing and spin-dependent transmission process across the domain wall makes $P$ extrinsically dependent on the local magnetization configuration, and also temperature. However, since $\beta$ is usually on the same order of $\alpha$, we can assume the factor $\frac{\alpha\beta+1}{1+\alpha^2}$ to be close to 1, and inversely derive the value of $P$ (Table S6). As can be seen, the derived $P$ values are in reasonable range, and will decrease as temperature becomes closer to $T_C$ as expected.

**Supplementary Table S6.** Derived transport spin polarization $P \approx 2\eta e M_S / g\mu_B$ as a function of temperature.

| | $T$ (K) | $P$ | | $T$ (K) | $P$ |
|---|---|---|---|---|---|
| Device 4 | 150 | 0.28 | Device 5 | 150 | 0.41 |
| | 160 | 0.25 | | 160 | 0.36 |
| | 170 | 0.18 | | 170 | 0.22 |

From the results above, we showed that both the field- and current-dependent DW mobility $\mu$ and $\eta$, and therefore the experimentally observed very high STT efficiency $\eta/\mu$, can be fully explained by the classical DW motion model after Walker breakdown. The STT efficiency is then

$$\frac{\Delta H}{\Delta J} \equiv \frac{\eta}{\mu} = \frac{g\mu_B}{2e\gamma} \cdot \frac{P}{M_S} \cdot \sqrt{\frac{K_u}{A}} \cdot \left(\beta + \frac{1}{\alpha}\right) \tag{5}$$

Eq (5) gives out essential ingredients for producing a large STT efficiency on a magnetic DW:

1. large uniaxial magnetic anisotropy $K_u$
2. small saturation magnetization $M_S$
3. large transport spin polarization $P$
4. small Gilbert damping factor $\alpha$
5. small exchange stiffness $A$

$Co_3Sn_2S_2$ happens to be a material that owns all the listed advantages. We note that the value of $\beta$ will not change STT efficiency much, in the normal case $\alpha \ll 1$. This means the experimentally observed high STT efficiency, as well as the high efficiency modulation of $H_C$, cannot be directly attributed to “Weyl fermions”.

Importantly, we clarify that here the direct impact of Weyl fermions on DW dynamics means the possible enhancement of the non-adiabatic STT coefficient $\beta$, which is a unique property of the low energy electronic excitations in a magnetic Weyl semimetal. However, the magnetic and transport parameters mentioned above, which directly lead to the high STT efficiency in $Co_3Sn_2S_2$, are also highly expected to stem from the special electronic band structure of a magnetic Weyl semimetal. We explain in detail below.

A large uniaxial magnetic anisotropy $K_u$ (for $Co_3Sn_2S_2$, the magnetocrystalline anisotropy), is more likely to exist in magnetic Weyl semimetals. The relative location of two Weyl nodes with opposite chirality, as well as the overall electronic band structure, of a magnetic Weyl semimetal, is known to be greatly dependent on the magnetization direction[32,33]. When the magnetization of $Co_3Sn_2S_2$ is perpendicular to the easy axis, the Weyl nodes become nodal rings (gaps close along the nodal ring). The total energy of the system is then much higher than the gapped case with Weyl nodes only. Also, the spin-orbit coupling in magnetic Weyl semimetals is usually strong. Both the factors help to improve $K_u$.

A small magnetization $M_S$ is not easily obtained in metallic ferromagnets, commonly around the value of 1000 emu/cm$^3$. By comparison, the $M_S$ of $Co_3Sn_2S_2$ is only 48.8 emu/cm$^3$ at 150 K and 72.4 emu/cm$^3$ at 4 K. According to the Stoner criterion, itinerant ferromagnetism is stable when $g(E_F)U>1$, where $g(E_F)$ is the density of states in the non-magnetic state, and $U$ the onsite Coulomb interaction. As a result, conventionally, to realize ferromagnetism, $g(E_F)$ must be high. Take Fe, Co, Ni as an example, the high $g(E_F)$ leads to large number of uncompensated $3d$ electrons, and therefore, a large $M_S$. Magnetic Weyl semimetals are in this context a very special category of itinerant ferromagnets, which has a minimum of $g(E_F)$, making a small $M_S$ possible. The origin of such a low DOS ferromagnetism is beyond the scope of this paper but has been discussed elsewhere[33].

The transport spin polarization $P$ is again, naturally expected to be larger in magnetic Weyl semimetals, than conventional metals. Because of the small Fermi surface and splitting of the two spin bands, it is more likely that only one of the spin bands crosses $E_F$, making the band structure half-metallic. This is exactly the case in $Co_3Sn_2S_2$, leading to large spin polarization (we mentioned briefly the extrinsic nature of $P$, so experimentally $P$ can never reach 100%). Although the values of $\alpha$ and $A$ are not as strongly connected to Weyl semimetals, the discussion on $K_u$, $M_S$ and $P$ elucidates the benefits of choosing magnetic Weyl semimetals as a candidate for power-efficient spintronic devices. We also notice the recent report[34] on large STT efficiency in another ferromagnetic Weyl semimetal $SrRuO_3$ (although still 1-2 orders smaller than the value reported here but already orders larger than conventional metallic systems). Taken the considerations together, the high STT efficiency is possibly a general property expected in PMA magnetic Weyl semimetals.

Another unusual feature we report is the low current density threshold to drive a DW. Compared with conventional ferromagnetic metals that all have threshold current density on the order of $10^7$-$10^8$ A/cm$^2$ (based on pure STT), the magnetic Weyl semimetals $Co_3Sn_2S_2$ and $SrRuO_3$ are the only two metallic systems we know to have significantly smaller threshold current densities. In $SrRuO_3$, the current density needed to drive the wall[35] was $5.3\times10^5$ A/cm$^2$ at $T = 140$ K and $5.8\times10^6$ A/cm$^2$ at $T = 40$ K. Interestingly, if this current threshold is caused by extrinsic pinning (because $SrRuO_3$ has very thin domain wall ~ 3 nm), we will reach a non-adiabatic coefficient $\beta_w \approx 0.5$, which will be too large if interpreted as due to spin relaxation[36].

We use analytical DW motion model and micromagnetic simulation to investigate the origin of the low threshold. The adiabatic condition when the spin polarized current crosses a DW is $\Delta \cdot k_F \gg 1$, where $\Delta = \sqrt{\mathrm{A/K_u}}$, and $k_\mathrm{F}$ is the Fermi wavevector. In $Co_3Sn_2S_2$, $\Delta \approx 1.3$ nm, and $k_F \approx 0.06$ Å$^{-1}$, giving $\Delta \cdot k_F \approx 0.8$, at the boundary of adiabatic and non-adiabatic conditions. Therefore, we need to discuss both the possibilities. The small $k_F$ that could only exist in Weyl semimetals suggests another unique advantage of the topological quantum magnet.

In the adiabatic limit, the intrinsic current density threshold can be calculated by[36]

$$j_c^{\mathrm{in}} = \frac{e v_c^2 M_S^2 K_\perp \Delta}{P Z^2 \gamma^2 \hbar^3} \tag{6}$$

Where $v_\mathrm{c}$ is the volume of unit cell, $K_\perp$ is the hard axis anisotropy (along nanowire width direction), and $Z = 9$ is the number of cobalt atoms in the unit cell. Because $K_\perp = 4\pi N_y M_S^2 \propto M_S^2$, the intrinsic current density threshold is $j_c^{in} \propto M_S^4$, decreasing dramatically as $M_S$ decreases. Plugging in the magnetic parameters of $Co_3Sn_2S_2$, we find $j_c^{in} \approx 1.6 \times 10^3$ A/cm$^2$ at 150 K, two orders smaller than the current density we can reach in experiment. This low threshold is understandable because our discussion in Supplementary Note 6 indicates that the true current threshold could possibly approach zero.

In the non-adiabatic limit, the current density threshold is extrinsically dependent on the pinning strength,

$$j_c^{\mathrm{ex}} = \frac{e v_c^2 M_S^2 \Delta}{P Z^2 \gamma^2 \hbar^3} \cdot \frac{\mu_B B_c Z}{\beta_w v_c} \tag{7}$$

Where $\beta_w = \beta_{\mathrm{ST}} + \beta_{\mathrm{wall}}$ is the total non-adiabatic coefficient of STT ($\beta_{\mathrm{ST}}$) and domain wall resistance ($\beta_{\mathrm{wall}}$), and $B_c$ the depinning field. With a large $\beta$ value, the current threshold can be significantly lowered. $B_c$ = 50 G (reported in $SrRuO_3$ at 140 K; we assume a similar value for $Co_3Sn_2S_2$) corresponds to $j_c^{ex} \approx 7.1 \times 10^3\ \mathrm{A/cm^2}$ when $\beta = 1$, and $j_c^{ex} \approx 7.1 \times 10^4\ \mathrm{A/cm^2}$ when $\beta = 0.1$ for $Co_3Sn_2S_2$ at 150 K. We can also calculate the depinning effective field, $B_c/j_c$, to be 7.04 kG $\mathrm{MA^{-1}\ cm^2}$, when $\beta = 1$. As can be seen from eq (7), in the non-adiabatic limit and extrinsic pinning case, small $M_s$, large $K_u$ and large $\beta$ still helps reducing the threshold current density down to at least two orders smaller than conventional metals.

All the above discussions are cross verified by micromagnetic simulation[37] ($\mathrm{mumax}^3$ program). Grid size of (256×16×1) and cell size of (0.1nm, 0.2nm, 0.2nm) were chosen to capture the thin domain wall structure and our nanowire geometry. We use $\alpha$=0.05, $P$=0.35, and magnetic parameters in Table S4 at 150 K for the simulation. The current- and field-driven DW velocities are shown in Fig. S12 and S13. Interestingly, we did not observe a well-defined current threshold, even in the case of $\beta = 0$, which also agrees with our experimental results. The extracted DW mobilities are summarized in Table S7. As can be seen, micromagnetic simulation gives identical results as the analytical model. Therefore, the experimental data can be well explained.

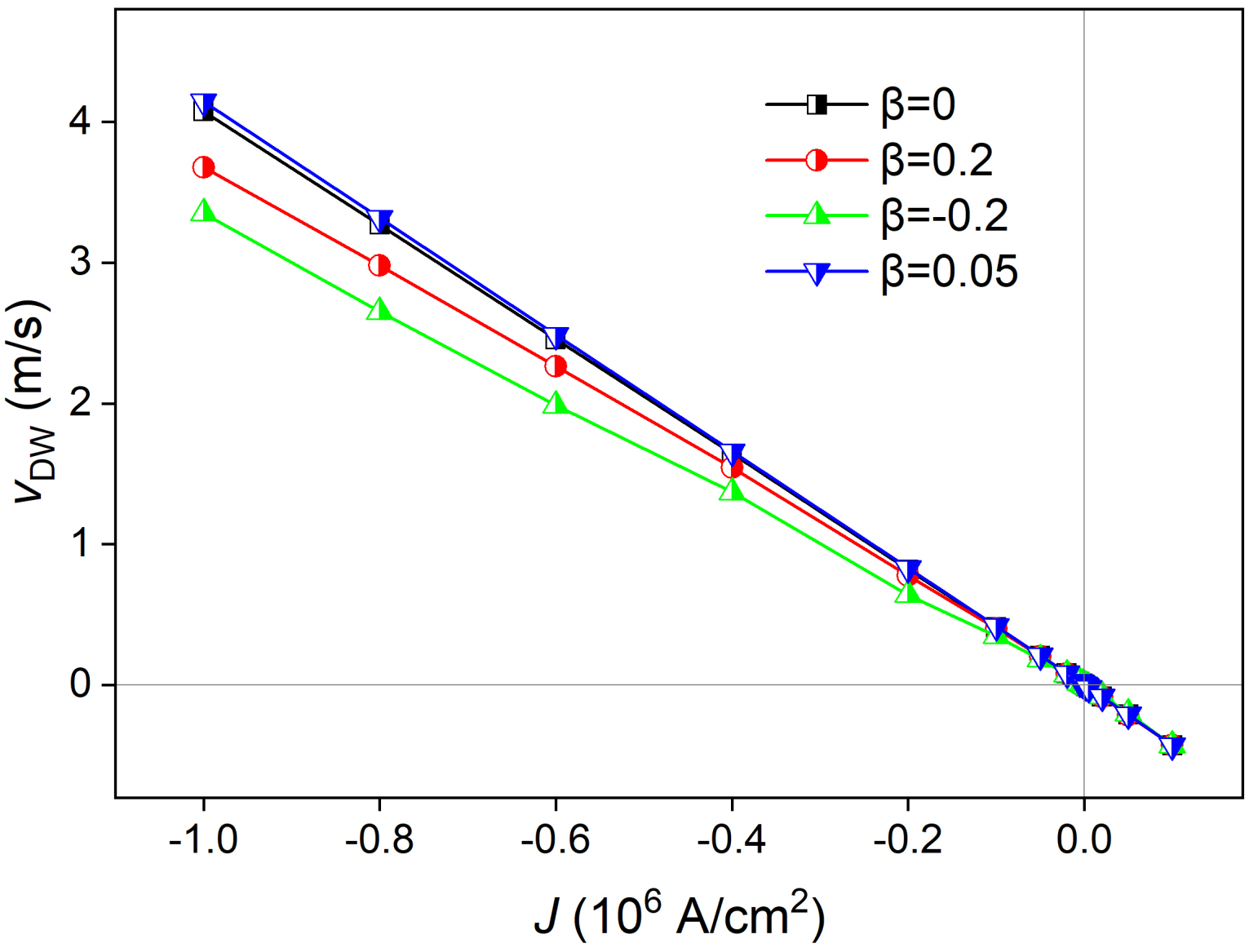

**Supplementary Fig. S12** DW velocity as a function of $J$ (simulation).

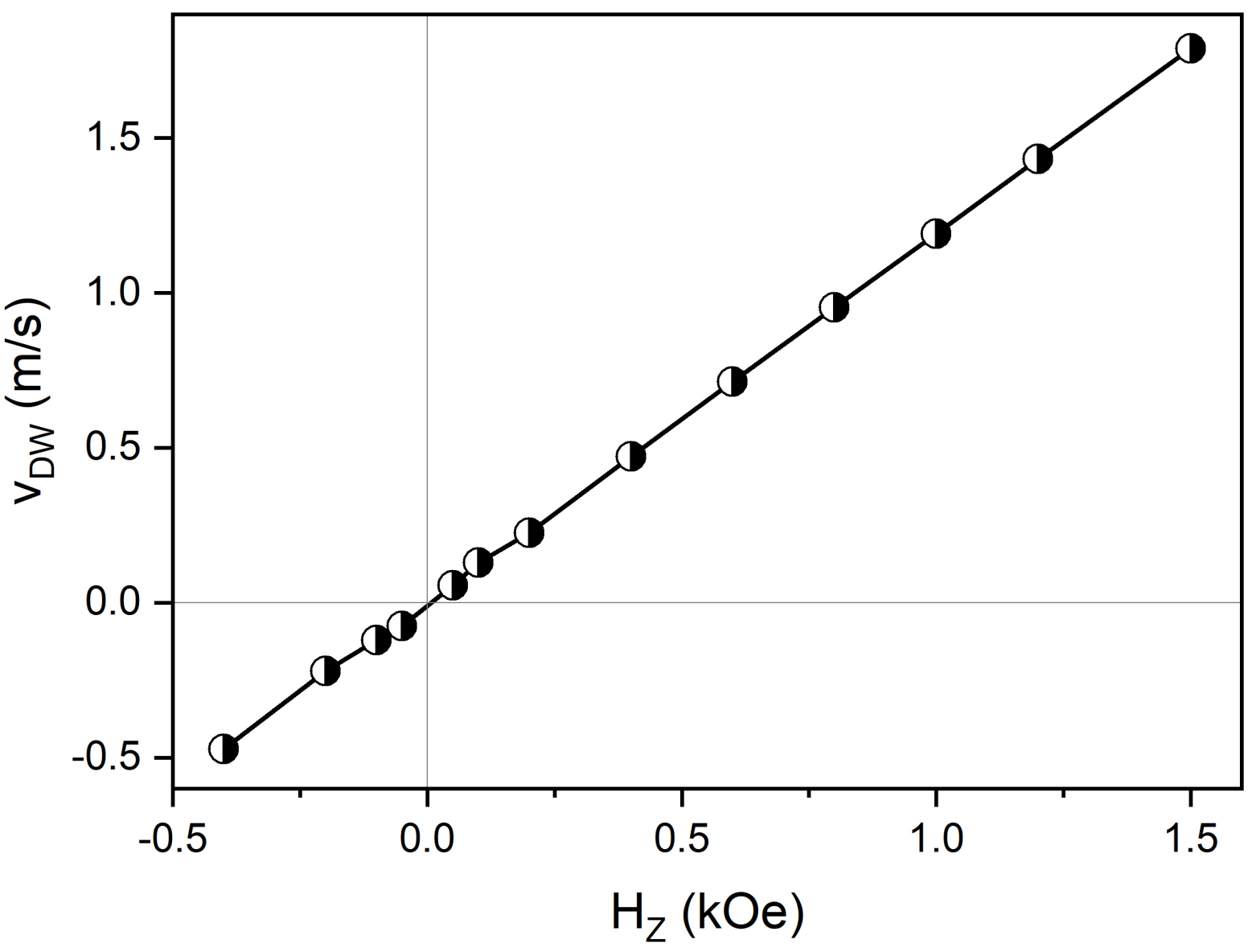

**Supplementary Fig. S13** DW velocity as a function of $H_z$ (simulation).

**Supplementary Table S7.** Comparison between experimental and simulated DW mobilities.

| | α | P | $\eta$ ($10^{-9}$ $m^3/C$) | $\mu$ ($10^{-3}$ m/(s·Oe)) |
|---|---|---|---|---|
| Device 4 | 0.057 | 0.28 | 0.328 | 1.357 |
| Device 5 | 0.036 | 0.41 | 0.483 | 0.861 |
| Simulation (β=0) | 0.05 | 0.35 | 0.407 | 1.191 |
| Analytical model | 0.05 | 0.35 | 0.411 | 1.193 |

At last, we feel necessary to comment that whether the observed features originate from "Weyl fermions" does not affect its technological application. Meanwhile, the exotic magnetic and transport parameters from special Weyl semimetal band structure should serve as a good guidance in the search of similar material candidates in the future.

## Supplementary References

1. Tanaka, M. *et al.* Topological Kagome Magnet $Co_3Sn_2S_2$ Thin Flakes with High Electron Mobility and Large Anomalous Hall Effect. *Nano Lett.* **20**, 7476–7481 (2020).
2. Yang, S.-Y. *et al.* Field-Modulated Anomalous Hall Conductivity and Planar Hall Effect in $Co_3Sn_2S_2$ Nanoflakes. *Nano Lett.* **20**, 7860–7867 (2020).
3. Fujiwara, K. *et al.* Ferromagnetic $Co_3Sn_2S_2$ thin films fabricated by co-sputtering. *Jpn. J. Appl. Phys.* **58**, 050912 (2019).
4. Ikeda, J. *et al.* Critical thickness for the emergence of Weyl features in Co3Sn2S2 thin films. *Commun Mater* **2**, 18 (2021).
5. Liu, E. *et al.* Giant anomalous Hall effect in a ferromagnetic kagome-lattice semimetal. *Nature physics* **14**, 1125–1131 (2018).
6. Howlader, S. *et al.* Spin-orbit driven spin depolarization in the ferromagnetic Weyl semimetal Co3Sn2S2. *arXiv preprint arXiv:1906.06557* (2019).

7. Wang, H. *et al.* Investigation of point-contact Andreev reflection on magnetic Weyl semimetal Co3Sn2S2. *Sci. China Phys. Mech. Astron.* **63**, 287411 (2020).
8. Kuz'min, M. D., Skokov, K. P., Diop, L. V. B., Radulov, I. A. & Gutfleisch, O. Exchange stiffness of ferromagnets. *Eur. Phys. J. Plus* **135**, 301 (2020).
9. Atxitia, U. *et al.* Micromagnetic modeling of laser-induced magnetization dynamics using the Landau-Lifshitz-Bloch equation. *Appl. Phys. Lett.* **91**, 232507 (2007).
10. Mougin, A., Cormier, M., Adam, J. P., Metaxas, P. J. & Ferré, J. Domain wall mobility, stability and Walker breakdown in magnetic nanowires. *Europhys. Lett.* **78**, 57007 (2007).
11. Thiaville, A. & Nakatani, Y. Micromagnetics of Domain Wall Dynamics in Soft Nanostrips. in *Nanomagnetism and Spintronics* 261–313 (Elsevier, 2014). doi:10.1016/B978-0-444-63279-1.00006-X.
12. Yin, J.-X. *et al.* Negative flat band magnetism in a spin–orbit-coupled correlated kagome magnet. *Nat. Phys.* **15**, 443–448 (2019).
13. Johansson, A., Henk, J. & Mertig, I. Edelstein effect in Weyl semimetals. *Phys. Rev. B* **97**, 085417 (2018).
14. Zhao, B. *et al.* Observation of charge to spin conversion in Weyl semimetal WTe 2 at room temperature. *Phys. Rev. Research* **2**, 013286 (2020).
15. Manchon, A. & Zhang, S. Theory of nonequilibrium intrinsic spin torque in a single nanomagnet. *Phys. Rev. B* **78**, 212405 (2008).
16. Hals, K. M. D. & Brataas, A. Phenomenology of current-induced spin-orbit torques. *Phys. Rev. B* **88**, 085423 (2013).
17. Chernyshov, A. *et al.* Evidence for reversible control of magnetization in a ferromagnetic material by means of spin–orbit magnetic field. *Nature Phys* **5**, 656–659 (2009).
18. Fan, Y. *et al.* Electric-field control of spin–orbit torque in a magnetically doped topological insulator. *Nature Nanotech* **11**, 352–359 (2016).
19. Jiang, M. *et al.* Efficient full spin–orbit torque switching in a single layer of a perpendicularly magnetized single-crystalline ferromagnet. *Nat Commun* **10**, 2590 (2019).
20. Ikeda, J. *et al.* Two-dimensionality of metallic surface conduction in Co3Sn2S2 thin films. *Commun Phys* **4**, 117 (2021).
21. Gushi, T. *et al.* Large Current Driven Domain Wall Mobility and Gate Tuning of Coercivity in Ferrimagnetic $Mn_4N$ Thin Films. *Nano Lett.* **19**, 8716–8723 (2019).
22. Suemasu, T., Vila, L. & Attané, J.-P. Present Status of Rare-earth Free Ferrimagnet $Mn_4N$ and Future Prospects of $Mn_4N$-based Compensated Ferrimagnets. *J. Phys. Soc. Jpn.* **90**, 081010 (2021).
23. Cullity, B. D. & Graham, C. D. *Introduction to magnetic materials*. (2015).
24. Kirilyuk, A., Ferré, J., Grolier, V., Jamet, J. P. & Renard, D. Magnetization reversal in ultrathin ferromagnetic films with perpendicular anisotropy. *Journal of Magnetism and Magnetic Materials* **171**, 45–63 (1997).
25. Le Gall, S. *et al.* Statistical study of domain-wall depinning induced by magnetic field and current in an epitaxial Co/Ni-based spin-valve wire. *Phys. Rev. B* **98**, 024401 (2018).
26. Narayanapillai, K., Qiu, X., Rhensius, J. & Yang, H. Thermally assisted domain wall nucleation in perpendicular anisotropy trilayer nanowires. *J. Phys. D: Appl. Phys.* **47**, 105005 (2014).
27. Phung, T. *et al.* Highly Efficient In-Line Magnetic Domain Wall Injector. *Nano Lett.* **15**,

835–841 (2015).
28. Li, Z. & Zhang, S. Thermally assisted magnetization reversal in the presence of a spin-transfer torque. *Phys. Rev. B* **69**, 134416 (2004).
29. Kurebayashi, D. & Nomura, K. Theory for spin torque in Weyl semimetal with magnetic texture. *Sci Rep* **9**, 5365 (2019).
30. Kurebayashi, D., Araki, Y. & Nomura, K. Microscopic Theory of Electrically Induced Spin Torques in Magnetic Weyl Semimetals. *J. Phys. Soc. Jpn.* **90**, 084702 (2021).
31. Beach, G. S. D., Tsoi, M. & Erskine, J. L. Current-induced domain wall motion. *Journal of Magnetism and Magnetic Materials* **320**, 1272–1281 (2008).
32. Ghimire, M. P. *et al.* Creating Weyl nodes and controlling their energy by magnetization rotation. *Phys. Rev. Research* **1**, 032044 (2019).
33. Ozawa, A. & Nomura, K. Two-Orbital Effective Model for Magnetic Weyl Semimetal in Kagome-Lattice Shandite. *J. Phys. Soc. Jpn.* **88**, 123703 (2019).
34. Yamanouchi, M. *et al.* Observation of topological Hall torque exerted on a domain wall in the ferromagnetic oxide $SrRuO_3$. *Sci. Adv.* **8**, eabl6192 (2022).
35. Feigenson, M., Reiner, J. W. & Klein, L. Efficient current-induced domain-wall displacement in SrRuO 3. *Physical review letters* **98**, 247204 (2007).
36. Tatara, G., Kohno, H. & Shibata, J. Microscopic approach to current-driven domain wall dynamics. *Physics Reports* **468**, 213–301 (2008).
37. Vansteenkiste, A. *et al.* The design and verification of MuMax3. *AIP Advances* **4**, 107133 (2014).